\documentclass[amsmath,amssymb,notitlepage, twocolumn,aps,pra,floatfix]{revtex4-2}
\usepackage[english]{babel}
\usepackage{xcolor}
\usepackage[colorlinks=true, allcolors = black]{hyperref}
\usepackage{url}
\usepackage{qcircuit}
\usepackage{subfigure}
\usepackage{graphicx}% Include figure files
\usepackage{dcolumn}% Align table columns on the decimal point
\usepackage{bm}% bold math
\usepackage{siunitx}
\usepackage{placeins}
\usepackage{comment}
\usepackage[T1]{fontenc}
\usepackage{textcomp}
\usepackage{xparse}
\usepackage{adjustbox}
\usepackage{multirow}
\usepackage{braket}
\usepackage{amsmath}
\usepackage{cleveref}
\usepackage{mathtools}

\DeclareMathOperator{\Tr}{Tr}
\newcommand{\abs}[1]{\lvert#1\rvert}

\AtBeginDocument{\let\oldcontentsline\contentsline}
\newcommand{\notoccontentsline}[4]{\oldcontentsline{#1}{}{}{#4}}
\newcommand{\droptocpage}{\addtocontents{toc}{\let\protect\contentsline\protect\notoccontentsline}}
\newcommand{\incltocpage}{\addtocontents{toc}{\let\protect\contentsline\protect\oldcontentsline}}

\begin{document}
\preprint{APS/123-QED}

\title{Quantum-network nodes with real-time noise mitigation using spectator qubits}

\author{S. J. H. Loenen$^{1,2}$}\thanks{These authors contributed equally to this work.}
\author{Y. Wang$^{1,3}$}\thanks{These authors contributed equally to this work.}
\author{N. Demetriou$^{1,2}$}
\author{C. E. Bradley$^{1}$}
% \author{M. Markham$^4$}
% \author{D. J. Twitchen$^4$}
\author{T. H. Taminiau$^{1,2}$}
\email{T.H.Taminiau@TUDelft.nl}

\affiliation{$^{1}$QuTech, Delft University of Technology, PO Box 5046, 2600 GA Delft, The Netherlands}
\affiliation{$^{2}$Kavli Institute of Nanoscience Delft, Delft University of Technology, PO Box 5046, 2600 GA Delft, The Netherlands}
\affiliation{$^{3}$3. Physikalisches Institut, ZAQuant, University of Stuttgart, Allmandring 13, 70569 Stuttgart, Germany}
% \affiliation{$^{4}$Element Six, Fermi Avenue, Harwell Oxford, Didcot, Oxfordshire, OX11 0QR, United Kingdom}

\date{\today}

\begin{abstract}
Quantum networks might enable quantum communication and distributed quantum computation. Solid-state defects are promising platforms for such networks, because they provide an optical interface for remote entanglement distribution and a nuclear-spin register to store and process quantum information. A key challenge towards larger networks is to improve the storage of previously generated entangled states during new entanglement generation. Here, we introduce a method that uses `spectator' qubits combined with real-time decision making and feedforward to mitigate dephasing of stored quantum states during remote entanglement sequences. We implement the protocol using a single NV center in diamond and demonstrate improved memory fidelity. Our results show that spectator qubits can improve quantum network memory using minimal overhead and naturally present resources, making them a promising addition for near-term testbeds for quantum networks.
\end{abstract}

\maketitle

\noindent\textbf{INTRODUCTION}\\
%\section{Introduction}
\indent Quantum networks hold promise for a multitude of applications, ranging from secure communication and quantum sensor networks to distributed quantum computation \cite{wehner_quantum_2018, jiang_distributed_2007, nickerson_topological_2013, nickerson_freely_2014,Wang2024SA}. A potential architecture consists of quantum nodes that each contain a qubit register that provides quantum processing power, including quantum memory, and an optical interface to establish entangled links between different nodes \cite{kimble_quantum_2008}. The NV center in diamond, like other solid-state defects in wide-bandgap semiconductors \cite{pompili_realization_2021,nguyen_quantum_2019,babin_fabrication_2022,awschalom_quantum_2018,Wang2023PhdThesis}, provides these elements. The electronic spin can be interfaced with photonic qubits so that entanglement can be established \cite{bernien_heralded_2013,humphreys_deterministic_2018}. In addition, the defect host material provides long-lived nuclear-spin qubits that can be used as robust quantum memories as well as processing qubits via their interaction with the electronic spin \cite{bradley_ten-qubit_2019}. 

Following pioneering work on two-node quantum networks in different platforms \cite{stockill_phase-tuned_2017,stephenson_high-rate_2020,humphreys_deterministic_2018,knaut2023}, currently the largest quantum network able to run quantum protocols, such as entanglement swapping and quantum teleportation, consists of three solid-state defect nodes \cite{pompili_realization_2021,hermans_qubit_2022}. To efficiently create larger networks, it is desirable to faithfully store previously generated entangled states while generating new entangled links. Nuclear-spin qubits provide a robust memory. However, in network demonstrations to date, their decoherence rate under entanglement generation is still larger than the rate at which new entangled states are generated \cite{pompili_realization_2021, Bone2024}, which is captured by an active link efficiency smaller than 1 \cite{bradley_robust_2022}. On the one hand, a substantial amount of work is dedicated to improving the entanglement generation rate by incorporating defects into cavities \cite{faraon_resonant_2011,ruf_resonant_2021,nguyen_quantum_2019,lukin_4h-silicon-carbide--insulator_2020}. On the other hand, the memory has been made more robust against entanglement generation attempts by optimisation of the entanglement generation process  \cite{Beukers2019MScThesis,pompili_realization_2021}, as well as by reducing the effective coupling of the memory to the noise \cite{reiserer_robust_2016,bradley_robust_2022,bartling_entanglement_2022}.

This work introduces a complementary approach to protect the quantum network memory, which is based on sensing the noise experienced by the memory qubit during entanglement generation using auxiliary `spectator' qubits \cite{majumder_real-time_2020,orrell_sensor-assisted_2021,singh_mid-circuit_2023,lingenfelter_surpassing_2023,gupta_integration_2020,youssry_noise_2023,song_optimized_2023,tonekaboni_greedy_2023}. The underpinning idea is that the spectator qubits sense the noise process of interest, so that real-time feedforward, in combination with knowledge of how noise on the spectators correlates to the noise on the memory qubit, enables mitigation of the loss of quantum information stored in that memory qubit.

A key difference with respect to other methods, such as decoherence-protected subspaces or quantum error correction \cite{reiserer_robust_2016,abobeih_fault-tolerant_2022}, is that the spectator qubit approach does not require the encoding of the memory qubit in an entangled state before the noise process acts, thus reducing overhead and the associated initial loss of fidelity. Instead, the decision to act on the information accumulated on the spectators can be made dynamically after the noise process has acted. In cases where little noise was created, for example because the remote entanglement generation succeeded early, the operations on the spectators can be omitted, thereby avoiding the associated overhead and dephasing. Conversely, if remote entanglement generation induced significant noise, the spectators are used and reduce the total dephasing. Related approaches with spectator qubits have previously been proposed for real-time system calibration in trapped-ion qubits \cite{majumder_real-time_2020}, for monitoring energy injection events in superconducting qubits \cite{orrell_sensor-assisted_2021} and inspired work on photonic spectator modes. Experimentally, spectator qubits have been implemented for mid-circuit error correction in dual-species cold atom arrays \cite{singh_mid-circuit_2023}.

\begin{figure*}[!ht]
\includegraphics[width=\textwidth]{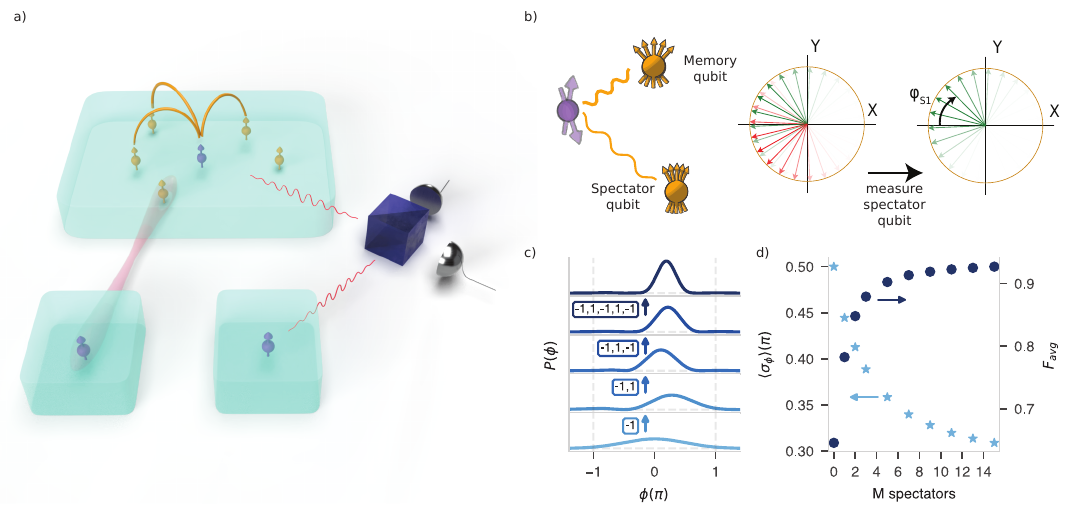}
\caption{\label{Figure1} \textbf{Concept: spectator qubits for network nodes.} \textbf{a)} Schematic of the spectator qubit protocol in a quantum network setting. Blue blocks indicate separate network nodes that each hold a communication qubit (purple) that is used as an optical interface. Additional qubits (only drawn in node 1) interact with the communication qubit and can serve different purposes, such as memory qubit (hold previously generated entangled state), idle qubit (store newly generated entangled state) or spectator qubits (provide information on memory qubit dephasing, indicated with yellow half circles connected to the communication qubit).  \textbf{b)} Example of correlated noise between the memory and spectator qubit. (left circle) The complete memory-qubit phase distribution before the spectator readout consists of the weighted sum of the memory-qubit phase distributions conditioned on a spectator qubit measurement outcome in orthogonal measurement bases in the XY plane (along the +Y axis, green, or the -Y axis, red). (right circle) Memory-qubit phase distribution after the spectator qubit is measured in the +Y axis. \textbf{c)} Narrowing of the memory-qubit phase probability density distribution for a specific readout syndrome of M spectators. Here we set $A_\perp=0$ and take $A_\parallel$ to be identical for all qubits ($g_i = 1)$. For M=0, the state of the memory (and each spectator) qubit is $\ket{\psi} = \frac{1}{2} \left[ \ket{0} + e^{-i\phi} \ket{1} \right]$, with $P_0(\phi) = \frac{1}{\sigma\sqrt{2\pi}}\exp\left(-\frac{1}{2}\frac{\phi^2}{\sigma^2}\right)$ and $\sigma = \frac{1}{2}\pi$. As an example, we consider the case with readout outcomes $m_i = 1$ for the even-numbered spectator qubits and $m_i = 0$ for the odd ones. In this calculation, gates and measurements are ideal. \textbf{d)} Standard deviation $\langle \sigma_\phi \rangle$ of the memory qubit phase distribution averaged over all possible spectator readout outcomes), and the corresponding memory-qubit fidelity $F_{avg}$. }
\end{figure*}

In this work, we investigate the spectator qubit approach in a quantum network setting, as demonstrated in Fig.~\ref{Figure1}. We demonstrate that the usage of multiple spectator qubits can extend the coherence time of a quantum memory during a control sequence that emulates remote entanglement generation. We use a NV-center as a quantum node, where the nuclear-spin register provides a quantum memory and multiple spectators. The electronic spin of the NV center serves as an optical interface to emulate the distribution of entanglement over different nodes, as well as to initialize, measure and control the nuclear spins. During the probabilistic entanglement generation sequence, the electron spin undergoes stochastic evolution, which, due to the always-on electron-nuclear hyperfine coupling, imposes spatially correlated dephasing on the memory and spectator qubits \cite{Blok2015,reiserer_robust_2016}.

Furthermore, we investigate different strategies for using the spectator qubits. First, we demonstrate that the overhead required to extract the spectator information imparts noise onto the system of interest and hence defines a trade-off as to when it is beneficial to access spectator qubits. Then we identify the measurement of the spectator qubits as one of the main noise sources, and mitigate this by employing a gate-based (without mid-circuit measurements) implementation. Based on these results, we show which spectator strategies to use given an entanglement success probability per entanglement generation trial. These results indicate that employing nuclear spins as spectator qubits provides a viable path to extend the memory coherence time when the noise experienced by the nuclear spins is dominantly correlated.\\

\noindent\textbf{BAYESIAN INFERENCE OF THE PHASE}\\
\indent We investigate the spectator qubit approach in a well-characterized NV system described by Abobeih et al \cite{abobeih_atomic-scale_2019}. The NV electronic spin couples to a register of nuclear spin qubits, of which the dynamics are described by

\begin{align}
    H_{n_i} = \, \omega_l I_z&\otimes \ket{0}\bra{0}_e + \label{eq:Hamiltonian_n_electronDependent} \\
             & \left[ (\omega_l - A_{\parallel_i})I_z + A_{\perp_i} I_x \right]\otimes\ket{1}\bra{1}_e, \nonumber
\end{align}

\noindent where \textquotedbl$i$\textquotedbl\ labels each single nuclear spin, $\omega_l$ is the nuclear-spin qubit Larmor frequency, \{$A_\parallel$, $A_\perp$\} the hyperfine coupling parameters to the electron spin qubit and $\ket{m_s}_e$ denotes the electron spin state ($\ket{0}$, $\ket{1}$ corresponding to the $m_s$ = 0, -1 projections respectively). For sequences with stochastic electron-spin evolution (such as investigated in this work), the hyperfine interaction introduces decoherence on the nuclear spin qubits.

To provide insight into the spectator qubit approach, as an example, we first consider the limit where $\frac{A_{\perp_i}}{\omega_l \pm A_{\parallel_i}}<<1$. In this case, the two electron-spin-conditioned nuclear-spin dynamics in Eq.~\eqref{eq:Hamiltonian_n_electronDependent} commute, resulting in decoherence that manifests purely as dephasing. Furthermore, the correlation of the phase evolution of different nuclear-spin qubits is then set by their parallel hyperfine components $A_{\parallel}$.

Measurements on the spectator qubits reveal these correlations and facilitate a Bayesian update (SI section \ref{sec:SI_Bayesian}) of the memory qubit's phase distribution according to the likelihood function:
\begin{equation}
    P(\phi|\mathcal{M}_n)\propto \prod_{i=1}^{n} \frac{1 + (-1)^{m_i} \cos(g_i\phi-\theta_{i})}{2}\,P_0(\phi).
    \label{eq:Bayesian}
\end{equation}

\noindent Here $P(\phi|\mathcal{M}_n)$ is the memory (or spectator) qubit phase distribution given a string $\mathcal{M}_n$ of $n$ spectator readout outcomes $m_i$, with $m_i \, \epsilon \, \{0,1\}$. $g_i = \frac{A_{\parallel_i}}{A_{\parallel_m}}$ is the ratio of the parallel hyperfine couplings of spectator $i$ ($A_{\parallel_i}$) and the memory qubit ($A_{\parallel_m}$), $\theta_i$ denotes the basis in the XY-plane of the Bloch sphere along which spectator $i$ is read out, and $P_0(\phi)$ is the memory qubit phase distribution prior to any spectator measurements. Fig.~\ref{Figure1}b provides a graphical representation of spectator-measurement induced phase narrowing of the memory qubit. 

Equation \eqref{eq:Bayesian} also gives the updated phase distribution of any not-yet-measured spectator qubits. Fig.~\ref{Figure1}c and d exemplify the narrowing of the memory qubit phase distribution, and corresponding improved fidelity, after measuring $M$ spectators, where each spectator is measured in a basis perpendicular to its predicted Bloch vector (SI section \ref{sec:RObasis_phi_c}). Note that, even in this idealized case, extending to a general analytical scaling for more spectator qubits is nontrivial, as it requires averaging equation \eqref{eq:Bayesian} over all spectator measurement syndromes, and each $\theta_i$ itself is a function of equation \eqref{eq:Bayesian} for $i-1$ spectator qubits.\\

\noindent\textbf{SPECTATOR QUBITS IN ENTANGLEMENT PROTOCOLS}

\begin{figure*}[!ht]
\includegraphics[width=\textwidth]{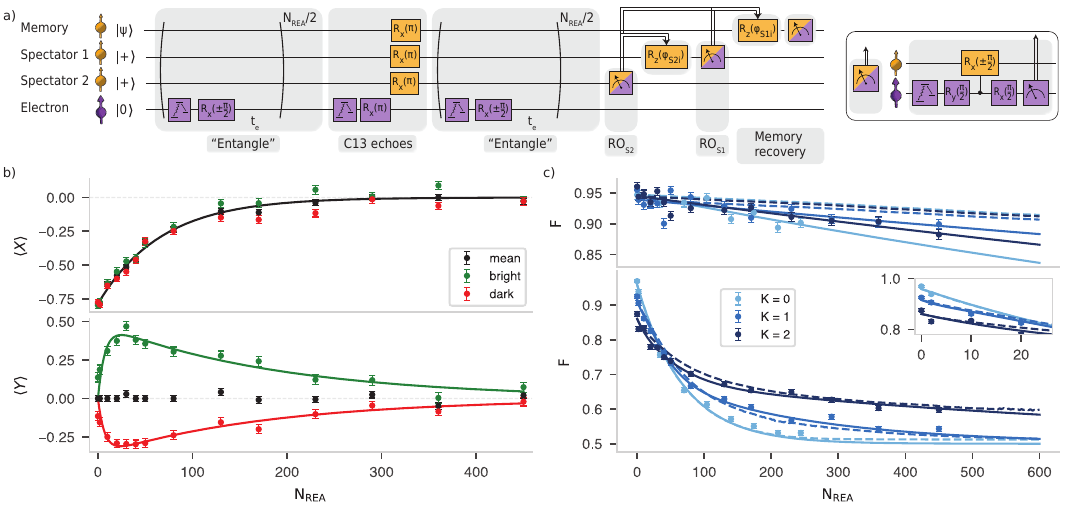}
\caption{\textbf{Spectator-based real-time noise mitigation.} \textbf{a)} Experimental sequence with $\mathrm{M=2}$ spectator qubits. After $\mathrm{N_{REA}}$ entanglement attempts, K spectators are read out (K=M=2 in the sequence shown). To maximize the information gain, the readout basis of each spectator (set by a $R_Z(\phi)$ rotation), is determined in real-time by the combined measurement syndrome of previously read-out spectators (see SI section \ref{sec:RObasis_phi_c}). \textbf{b)} Experimental data of the evolution of the memory qubit $\langle \mathrm{X} \rangle$ and $\langle \mathrm{Y} \rangle$ after running the protocol in (a) with the memory qubit initialized in the $\mathrm{\ket{+X}}$ state and a spectator with $g=1.49$. Data is conditioned on obtaining a bright (green) or dark (red) readout on the first spectator. Black data is unconditioned. \textbf{c)} Fidelity $F$ of the final state after the protocol in (a). (top) Memory qubit initialized in $\ket{0}$. (bottom) Average over data with memory-qubit initial states in $\mathrm{\ket{+X}}$ and $\mathrm{\ket{+Y}}$ for different number of spectator qubits initialized and read out. For low $\mathrm{N_{REA}}$, the measurement of the spectators imparts more dephasing noise on the memory qubit than the amount of dephasing that is compensated (see inset). Solid lines are fits and dashed lines are simulations (see SI sections \ref{sec:SI_1Spec_analytical} and \ref{sec:Simulation_Derivation}).}
\label{Figure2}  
\end{figure*}

%\section{Spectator qubits in remote entangling protocols}
We apply this approach in an experimental setting where we emulate a sequence of $\mathrm{N_{REA}}$ remote entanglement attempts after initializing M spectator qubits in the X-basis, see Fig.~\ref{Figure2}a with M=2. Here, a single entanglement attempt consists of a stochastic electron-spin reset,the preparation of a superposition state on the electron-spin qubit and a waiting time $t_e$ that, for example, allows the heralding signals of entanglement to be received before the next electron-spin reset. Note that we omit the optical $\pi$-pulse to generate spin-photon entanglement for simplicity and that various other sequences and protocols are possible, including adding extra pulses on the electron spin \cite{humphreys_deterministic_2018,pompili_realization_2021,hermans_entangling_2023}. Our sequence provides an example to illustrate the principle of spectator qubits; the exact performance will depend the parameters of a given system and sequence, and the resulting noise correlations. To rephase quasi-static noise, we implement a nuclear-spin echo after half the entanglement attempts. 

The electron-spin state during the reset depends on the optical cycling of the electronic spin, where stochastic decays occur from the excited state directly, or through the meta-stable state, to the ground state \cite{awschalom_quantum_2018,Kalb2018Dephasing}. In addition, the reset projects the superposition state. Over many entanglement attempts, these processes generate dephasing on the nuclear-spin qubits, which is correlated by the nuclear-spin hyperfine parameters.  

In an actual setting, upon completion of all $\mathrm{N_{REA}}$ attempts, the state of the electronic spin would be transferred to an idle nuclear-spin qubit to free up the electron spin for subsequent spectator qubit readouts. As we focus on the dephasing process, and do not generate an entangled state between different nodes, such a step is omitted in this work. Therefore, at the end of the emulated entanglement generation sequence, we reset the electron spin and read out K $\leq$ M spectators sequentially via the electron-spin qubit. Metrics for well-performing spectator qubits are good gate fidelities, small $A_\perp$ (reducing non-commutivity) and a range of $g$-values depending on the expected dephasing (SI section \ref{sec:SI_1Spec_analytical}). Note that we can decide in real-time how many spectators we read out. This allows one to only measure a spectator at the end of the entanglement generation process if this will improve the coherence of the memory qubit, which can be calibrated in advance (see discussion below and Fig.~\ref{Figure2}c). Each spectator readout basis is set in real time based on the measurement outcomes of previously measured spectators. 

We now experimentally demonstrate spectator-qubit-induced phase narrowing. First, we perform an experiment using two nuclear spins labelled C0 and C1 with respective hyperfine parameters $\{A_\parallel, A_\perp\}$ given by $\{24.4, 24.8\} \, \mathrm{and} \, \{-36.3, 26.6\}$ kHz \cite{abobeih_atomic-scale_2019}. All data is taken with a magnetic field of 404 G along the NV symmetry axis. We execute the sequence in Fig.~\ref{Figure2}a with C0 as the memory qubit (initialized in $\mathrm{\ket{+X}}$, $\mathrm{\ket{+Y}}$ or $\ket{0}$) and C1 as spectator 1 (always initialized in $\mathrm{\ket{+X}}$) (`spectator 2' is not used). The dephasing on both qubits is correlated via their hyperfine parameters ($A_{\parallel}$) and increases for increasing $\mathrm{N_{REA}}$ (longer stochastic electron evolution). At the end of the sequence, we measure the memory qubit (<X>, <Y>, Fig. 2(b)). Conditioned on the spectator qubit read-out in the -Y basis (labeled as `bright' or `dark', for electron spin states $m_s=0$ and $m_s=-1$, respectively), we observe a projection of the memory qubit either towards the +Y or -Y basis, demonstrating the correlation. 

We now include a second spectator qubit C2 ($\{A_\parallel, A_\perp\} = \{20.6,41.5\}$ kHz). Correspondingly, the g-values of the spectator qubits are $g_{C1} = -1.49$ and $g_{C2} = 0.84$. The readout axis of the second spectator qubit can now be calibrated, given the read-out syndrome on the first spectator, to be perpendicular to it's predicted Bloch vector. To quantify the corresponding reduction of dephasing on the memory qubit, we consider the fidelity associated with the remaining memory-qubit coherence: $\mathrm{F} = \frac{1}{2}\sqrt{\langle X \rangle^2 + \langle Y \rangle^2 + \langle Z \rangle^2} + \frac{1}{2}$, as shown in Fig.~\ref{Figure2}c. 

For larger $\mathrm{N_{REA}}$ (increased dephasing), we observe a marked improvement in the final fidelity when using additional spectator qubits. Conversely, for small $\mathrm{N_{REA}}$ (i.e., no or limited dephasing), adding spectators reduces the fidelity due to the additional dephasing introduced by the operations on the spectators.

Because phase is a $2\pi$ cyclic variable, correlation between the spectator and memory qubit phase is a non-trivial function of the $g$-value (see SI section \ref{sec:SI_1Spec_analytical} and \ref{sec:Simulation_Derivation}). This explains the more significant improvement of spectator 2 compared to spectator 1 for large $\mathrm{N_{REA}}$, where a $g$-value closer to 1 is more optimal. Memory-qubit eigenstates along the Z-axis show a limited loss of fidelity (bit flips), confirming that the dominant process is dephasing. The spectator qubits approach implemented here is tailored to dephasing noise and does not protect the Z-basis states. We confirm our understanding by comparing the experimental data set with a simulation that models the sequence shown in Fig.~\ref{Figure2}a. See SI section \ref{sec:Simulation_Derivation} for a detailed discussion on the simulation. 

Imperfections in the spectator-qubit control and readout reduce the overall performance through two distinct mechanisms. First, a faulty spectator measurement outcome (e.g. an initialization or readout error) leads to an incorrect phase update of the memory qubit, reducing the net gain from the spectator qubit. Second, stochastic electron spin flips during the spectator readout lead to an additional unknown phase on the nuclear-spin qubits, causing additional dephasing (SI section \ref{sec:Measurement_Dephasing}). Hence, measuring a spectator is only desired if the net information gain outweighs the additional measurement-induced dephasing. While the additional dephasing is independent of the amount of entanglement attempts $\mathrm{N_{REA}}$ executed, the information gain from spectator measurements depends on $\mathrm{N_{REA}}$. Consequently, the optimal strategy related to which spectators to read out depends on $\mathrm{N_{REA}}$, which is demonstrated in Fig.~\ref{Figure2}c. For low $\mathrm{N_{REA}}$, see inset, the strategies involving spectator qubits provide a lower memory fidelity compared to those that do not use spectators. 

In the following, we present a gate-based  implementation that bypasses the need for readout of spectator qubits. \\

\noindent\textbf{GATE-BASED IMPLEMENTATION}\\
%\section{Gate-based spectator qubit implementation}
\indent Unlike previously considered applications of spectator qubits \cite{majumder_real-time_2020,orrell_sensor-assisted_2021,singh_mid-circuit_2023,lingenfelter_surpassing_2023}, our system naturally provides controlled two-qubit interactions between the spectator/memory qubits and the source of dephasing (the electron-spin qubit). This enables us to efficiently replace the measurement-based scheme with a gate-based scheme, by substituting the measurement and classical feedforward by an equivalent two-qubit gate and qubit reset. We now explain the gate-based implementation in detail.

\begin{figure*}[ht]
\includegraphics[width=\textwidth]{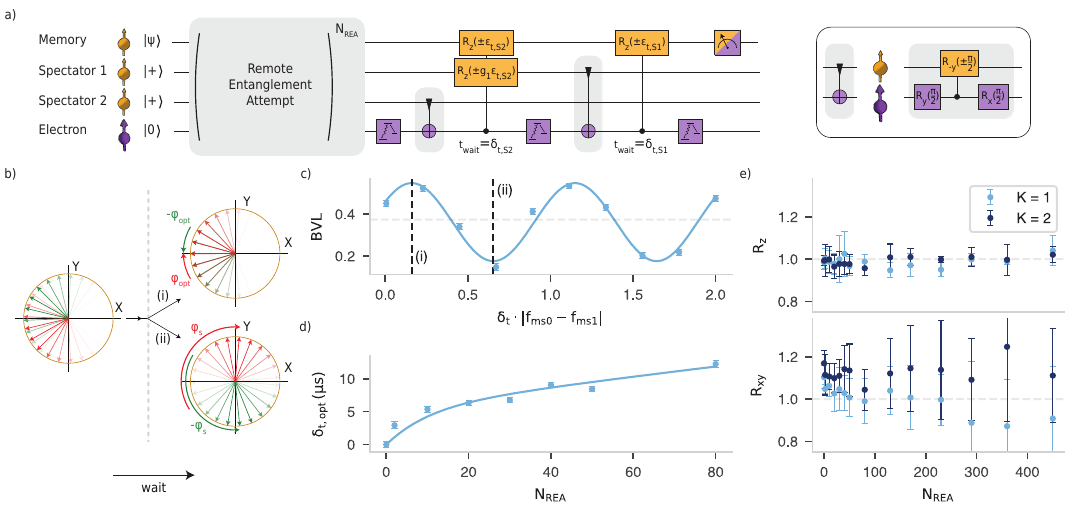}
\caption{\label{Figure3} \textbf{Gate-based spectator implementation.} \textbf{a)} Experimental sequence. Before memory retrieval, the state of a spectator qubit is mapped to the electron-spin qubit and a waiting time is applied that implements an electron-controlled $R_Z(\theta)$ operation on all nuclear spin qubits (SI section \ref{sec:SI_CZ}). Since the electron state is correlated with the phase of a spectator qubit, all nuclear spins that experienced correlated noise partly rephase if the interaction time is correctly set, see (b). The use of multiple spectators provides an accumulated rephasing. \textbf{b)} (left) Memory qubit phase distribution after mapping the Y-basis expectation value of a spectator qubit to the electron-spin Z-basis (|0>: red, |1>: green). (right) Upon applying a waiting time $\delta_{t}$, the two distributions acquire electron-spin dependent conditional phase. By timing optimally, rephasing occurs (top), whereas sub-optimal timing induces further dephasing (bottom). \textbf{c)} Experiment with $\mathrm{N_{REA}}$ = 30, C0 as the memory qubit and one spectator qubit (C1). The memory qubit Bloch vector length (BVL) with sinusoidal fit as a function of the waiting time after the nuclear-spin-conditioned electron rotation, (i) and (ii) correspond to the optimal and suboptimal rephasing times (see b). \textbf{d)} The optimal rephasing time $\delta_{t,opt}$ as a function of $\mathrm{N_{REA}}$ (SI section \ref{sec:SI_fitparameters} for fit). \textbf{e)} Quantification of the improvement of the gate-based implementation over the measurement-based implementation. Data is averaged over all permutations of C0, C1 and C2 acting as memory and spectator qubits. We plot the ratio of the memory-qubit Bloch vector length for the memory qubit initialized in $\ket{0}$ ($\mathrm{R_z}$, top) and averaged over the memory-qubit initial states $\mathrm{\ket{+X}}$ and $\mathrm{\ket{+Y}}$ ($\mathrm{R_{xy}}$, bottom). The vertical bars represent the standard deviation of the underlying distribution from which $\mathrm{R_z}$ and $\mathrm{R_{xy}}$ are calculated (not the statistical error on the mean value). No effect is observed in the $\mathrm{R_z}$, while a considerable improvement is observed in $\mathrm{R_{xy}}$ for the gate-based scheme.}
\end{figure*}

After $\mathrm{N_{REA}}$ repetitions of the remote entanglement sequence, we prepare the electron-spin qubit into an eigenstate. The goal is to (partly) revert the (unknown) phases picked up by the nuclear-spin qubits. To achieve this, we take two steps. First, we implement a spectator-controlled electron bit flip that correlates a $+\frac{\pi}{2}$ ($-\frac{\pi}{2}$) spectator phase (in the nuclear spin rotating frame with frequency $f_r = \frac{1}{2} (f_{\ket{0}_e} + f_{\ket{1}_e})$, the average over the two electron-spin-dependent nuclear-spin frequencies) with the electron state that induces a negative (positive) phase on that spectator (right panel of Fig.~\ref{Figure3}a). Second, we apply a waiting time $\delta_t$, which creates an electron-controlled phase rotation on all nuclear spins. 

Because the imparted phases are set by the same couplings ($A_{\parallel i}$) as those that created the dephasing, all nuclear spins can be refocused simultaneously. The efficiency of the phase reversal depends on correctly timing $\delta_t$ (Fig.~\ref{Figure3}b and SI section VIII). A too long $\delta_t$ overcompensates the obtained unknown nuclear-spin qubit phase and therefore effectively imparts additional dephasing.

We experimentally demonstrate this in Fig.~\ref{Figure3}c, where we plot the Bloch vector length (BVL) given by $\sqrt{\langle X \rangle^2 + \langle Y \rangle^2 + \langle Z \rangle^2}$ for different $\delta_t$. The best waiting time $\delta_{t,opt}$ corresponds to the first maximum, because only at this time all nuclear-spin qubits optimally rephase simultaneously. Timings of subsequent maxima, which are possible because of the $2\pi$ cyclicity of the phase, depend on the specific hyperfine couplings of individual nuclear spins (spectators as well as memory qubits). More entanglement attempts (larger $\mathrm{N_{REA}}$) will induce more stochastic dephasing and correspondingly require a longer waiting time $\delta_t$ to achieve optimal phase reversal (Fig.~\ref{Figure3}d). A challenge arises for nuclear-spin states that did not undergo the dephasing process during entanglement generation. For example, the nuclear spin to which a potential heralded entangled state is swapped actually dephases during the waiting time $\delta_t$. However, because the electron-spin state is now controlled rather than stochastically fluctuating, this additional dephasing can be resolved through an echo pulse at $\frac{1}{2}\delta_t$ on those nuclear spins.

We now implement the gate-based approach in the sequence of Fig.~\ref{Figure3}a, where we have calibrated $\delta_{t,S2}$ and $\delta_{t,S1}$ for both spectators, and compare this to the measurement-based implementation of Fig.~\ref{Figure2}. We compare the two approaches by implementing the two schemes for different permutations of C0, C1 and C2 (memory or spectator qubits, see SI section \ref{sec:SI_alternative_spectators} for the complete dataset).

For each permutation, we measure the ratio R of the gate-based BVL over the measurement-based BVL after $\mathrm{N_{REA}}$ entanglement generation sequences and with K spectator qubits. Subsequently, for different initial states of the memory qubit, we plot the average R obtained over the different permutations, together with the standard deviation of the underlying distribution (Fig.~\ref{Figure3}d). When the memory qubit is initialized in an eigenstate, the ratio ($R_z$) remains close to 1. In contrast, when initialized in $\mathrm{\ket{+X}}$ or $\mathrm{\ket{+Y}}$, the ratio ($R_{XY}$ is larger than 1 (for most, but not all permutations). These results confirm that the gate-based implementation provides an improvement over the measurement-based approach.\\

\noindent\textbf{OPTIMAL STRATEGY FOR DIFFERENT SUCCESS PROBABILITIES}\\
\indent We now analyze what the optimal number of spectators is for a given probabilistic entanglement generation process.  We consider a probability $p$ to successfully generate entanglement in each attempt. The likelihood for success at attempt $\mathrm{N_{REA}}$ is then given by $\mathrm{P(\mathrm{N_{REA}})} = (1-p)^{\mathrm{N_{REA}}-1}\cdot p$. We analyze the use of spectator qubits as a function of $p$ by considering what the expected fidelity ($\mathrm{\overline{F}} = \sum_{\mathrm{N_{REA}}=1}^\infty \mathrm{F(\mathrm{N_{REA}})} \cdot \mathrm{P(\mathrm{N_{REA}})}$) of the memory qubit is with respect to its initial state. $\mathrm{F(\mathrm{N_{REA}})}$ is the memory qubit fidelity at attempt $\mathrm{N_{REA}}$.

\begin{figure}[ht]
\includegraphics[width=1 \columnwidth]{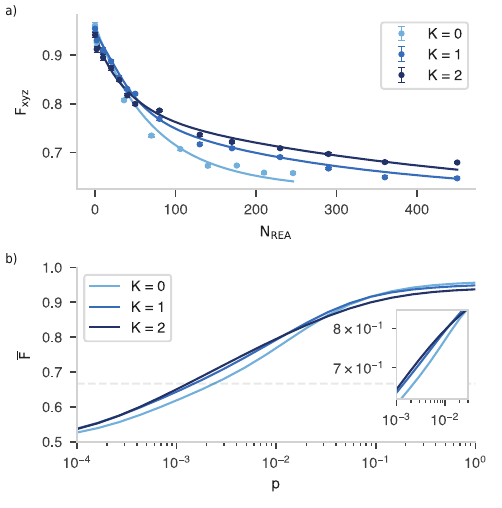}
\caption{\label{Figure4} \textbf{Memory fidelity $\mathrm{F}$ for different entanglement generation success probabilities using spectator qubits.} \textbf{a)} Memory fidelity $\mathrm{F}$ after $\mathrm{N_{REA}}$ entanglement generation attempts in the gate-based implementation using K spectators, averaged over the memory initial states $\ket{0}$, $\mathrm{\ket{+X}}$ and $\mathrm{\ket{+Y}}$. \textbf{b)}  Using the fits (solid lines in a, see SI section \ref{sec:SI_1Spec_analytical}), we plot  the expected average memory-qubit fidelity ($\mathrm{\overline{F}}$, main text) for different number (K) of spectators used.}
\end{figure}

We investigate and execute the gate-based implementation with the spectator and memory qubits identical to the ones used in Fig.~\ref{Figure2}c. We consider the situation where, after completion of $\mathrm{N_{REA}}$ entanglement sequences, we use the phase-information contained in each spectator that was initialized. Furthermore, we analyze the cases with 0, 1 or 2 spectators initialized. Fig.~\ref{Figure4}a shows the memory fidelity $\mathrm{F}$ averaged over datasets with the memory qubit initialized in the X-, Y- and Z-basis. In addition, we have analyzed datasets with the memory qubit initialized in the -Z-basis, which showed no significant difference from data with the memory qubit initialized in the +Z-basis. We fit the data using the analytical result for using a single spectator (SI section \ref{sec:SI_1Spec_analytical}), where we fix the state preparation and measurement (SPAM) error such that the fit overlaps with the data at $\mathrm{N_{REA}} = 0$. Using these fits, we interpolate the data and calculate $\mathrm{\overline{F}}$.

For low entanglement success probability p, there is a relatively high likelihood to generate entanglement at large $\mathrm{N_{REA}}$, where using spectators considerably outperforms not using spectators (Fig.~\ref{Figure2}c). This explains the regime in Fig.~\ref{Figure4}b where using spectators (K$\neq$0) outperforms using no spectators. On the contrary, for p $\approx$ 1, low numbers of $\mathrm{N_{REA}}$, where little correlated noise is built up, dominate $\mathrm{\overline{F}}$. Therefore, at low $\mathrm{N_{REA}}$, using spectators is not expected to provide a significant improvement over not using spectators. Rather, spectator-induced overhead can lower $\mathrm{\overline{F}}$ at low $\mathrm{N_{REA}}$.

A further optimisation strategy, which is not implemented here, is to decide in real time, for each spectator and based on $\mathrm{N_{REA}}$ until success for that instance, to use the information contained or not, e.g. to apply the measurement- or gate-based scheme. As the spectator qubit initialization comes at little costs (SI Fig.~\ref{Fig:SI_readout_dephasing}), an optimal strategy could be to always initialise multiple spectators and then decide which ones to use based on the number of sequence repetitions until success. Note that this would approximately follow the highest values of the curves in Fig.~\ref{Figure4}b (up to a potential initialization cost). Besides a longer required system initialization time, the cost of using more spectator qubits is an increased one-time (numerical or experimental) calibration cost of spectator read-out angles (measurement-based) or electron wait times (gate-based). Finally, we note that the spectator approach enables using qubits with limited fidelities, which might not be suitable to serve other purposes, as one can always decide at the end of the sequence to not use a spectator unless the dephasing was strong enough that it will improve memory qubit fidelity. \\

%\FloatBarrier

\noindent\textbf{CONCLUSION AND OUTLOOK}\\
%\section{Conclusion and outlook}
In this work, we used an NV center in diamond to demonstrate a method that uses multiple spectator qubits to mitigate the dephasing of quantum states stored in a quantum network node under an (emulated) entanglement process. This approach relies on the sensing of correlated noise along one axis, so that measurements performed on the spectator qubits reveal partial information of the noise imposed on the memory qubit. For other types of systems and entanglement sequences, such as those including echo-pulses on the electron spin \cite{bernien_heralded_2013,pompili_realization_2021, stolk_metropolitan-scale_2024}, the type of noise and its degree of correlations can vary, increasing or decreasing the effectiveness of spectators. 

Advantages of the spectator approach are: (1) it can be combined with other methods; (2) it imparts little intrinsic overhead or fidelity loss in preparation, as entanglement between the spectator qubits and the memory qubit is not required; (3) it can make use of qubits with limited control fidelity, which might have no other use, for example additional nuclear spins surrounding an NV center or other defect centers; and (4) the decision to apply the gates and measurements needed to use the spectator information can be deferred until after the noise has acted. This allows a pre-characterized trade-off between information gain and added disturbance to ensure that each spectator is used only when it improves the final memory fidelity. A potential use-case is in protocols where the quantum state of the memory qubit is teleported away from the node \cite{hermans_qubit_2022}, allowing the spectator qubits to be measured afterwards without causing any disturbance. 

In conclusion, we show that spectator qubits can be used in various situations to reduce the dephasing of a quantum-network memory, potentially using already present resources at limited cost. Given that the process itself does not add considerable dephasing, then it is always preferable to obtain information about the noise. This makes spectator qubits a promising avenue for near-term quantum network testbeds, where even modest fidelity enhancements might unlock new capabilities.  \\

\noindent\textbf{Author contributions}\\
SJHL, YW and THT devised the project and experiments. SJHL took and analyzed all experimental data with input from THT and YW. YW conceived the original idea with proof-of-principle theory calculations and simulations. SJHL developed the theoretical framework to understand and fit the experimental data. ND established the simulation framework applied to the experimental data, on which SJHL expanded. CEB assisted in analyzing data and taking preliminary data. SJHL wrote the manuscript with input from all authors. THT supervised the project.\\

\noindent\textbf{Acknowledgements}\\
We acknowledge M. Markham and D.J. Twitchen for the supplying the sample. We want to thank B. Terhal for useful discussions. This publication is part of the QuTech NWO funding 2020-2024 — Part I ``Fundamental Research'' with project number 601.QT.001-1, which is financed by the Dutch Research Council (NWO). This work is supported by German Federal Ministry of Education and Research (BMBF) for the project QECHQS with Grant No. 16KIS1590K. This work was supported by the Netherlands Organisation for Scientific Research (NWO/OCW) through a Vidi grant (680-47-552). This project has received funding from the European Research Council (ERC) under the European Union’s Horizon 2020 research and innovation programme (grant agreement No. 852410).\\

\noindent\textbf{Data Availability}\\
Data and code to reproduce the figures in this manuscript are available via DOI 10.4121/3929b39b-06e7-45c3-967a-6801e2ea3ae1. 

\droptocpage
\bibliography{Spectator_Paper/Spectator_Paper}
\incltocpage

\clearpage

% ###################################################
%%%%%%  Supplementary Information %%%%%%%%%%%%%%%%%%
% ################################################
\widetext
\begin{center}
\textbf{\large Supplementary Information for \\\textquotedbl Quantum-network nodes with real-time noise mitigation using spectator qubits\textquotedbl}
\end{center}

\renewcommand{\figurename}{\textbf{Supplementary Figure}}
\renewcommand{\tablename}{\textbf{Supplementary Table}}

\setcounter{equation}{0}
\setcounter{figure}{0}
\setcounter{table}{0}
\setcounter{page}{1}
\makeatletter
\renewcommand{\theequation}{S\arabic{equation}}
\renewcommand{\bibnumfmt}[1]{[S#1]}
\renewcommand{ \citenumfont}[1]{S#1}

\tableofcontents{}
\clearpage

\section{SUPPLEMENTARY NOTE: Correlation of nuclear spin evolution}
\label{sec:SI_Aperp}
In this section, we demonstrate that for $A_\perp = 0$ and the electron spin as the only fluctuator, the memory and spectator qubit evolutions are completely correlated by their parallel hyperfine component $A_\parallel$. Following Eq.~(1) in the main text, we can write the Hamiltonians of nuclear spin $i$ in the case the electron is in $\ket{0}$ and $\ket{1}$:

\begin{align}
    H_{0_i} &= \omega_l I_z\\
    H_{1_i} &= (\omega_l + A_{\parallel_i}) I_z + A_{\perp_i} I_x.
\end{align}

\noindent For stochastic electron evolution events, nuclear spin $i$ experiences $H_{0_i}$ and $H_{1_i}$ alternately for random times. For $M$ stochastic electron spin states, where the time spent in each state is random, the total nuclear spin evolution then equals

\begin{align}
    U_{i}(T) = \prod_j^M \exp{\left(-i H_{{\alpha_{M-j}}_i} t_j \right)},
    \label{eq:n_evo}
\end{align}

\noindent where $\alpha_{M-j} \in \{0,1\}$ and $\sum t_j = T$. If $A_\perp = 0$, then $H_{0_i}$ and $H_{1_i}$ commute. Therefore, following the Baker-Campbell-Hausdorff lemma, we can express as

\begin{align}
    U_{i}(T) = \exp{\left(-i \right[ \omega_l T + A_{\parallel_i} T_1\left] I_z \right)},
    \label{eq:n_evo_zero_Aperp}
\end{align}

\noindent where $T_1 \leq T$ is the total time the electron spent in the $\ket{1}$ state. The time ordering of when the electron is in state $\ket{0}$ and $\ket{1}$ is thus not important. From this equation we see that for two nuclear spins which only differ by their $A_\parallel$, the phase of two nuclear spins is completely correlated. Namely, relative to $\omega_l T$, if the phase of nuclear spin 1 is given by $\phi_1 = A_{\parallel_1}T_1$, then the phase of nuclear spin 2 equals $\frac{A_{\parallel_2}}{A_{\parallel_1}} \phi_1$.\\

In case $A_\perp \neq 0$, then $H_{0_i}$ and $H_{1_i}$ do not commute and the time ordering of when the electron is in the state $\ket{0}$ and the state $\ket{1}$ thus matters. Therefore, we cannot generally write equation \eqref{eq:n_evo} in a simple form as equation \eqref{eq:n_evo_zero_Aperp} and express a simple general relation between the evolution of different nuclear spins. For high magnetic fields however, the precession axis of $H_{0_i}$ and $H_{1_i}$ asymptotically overlap and hence equation \eqref{eq:n_evo_zero_Aperp} is retrieved. In our experiments the typical $A_\perp \approx \SI{20}{\kilo\hertz}$ which for our magnetic field of $\sim$403 Gauss gives a small tilt of the precession axis of $\sim$2.7 degrees. The simulations in figure \ref{Figure2}d, that include non-zero $A_\perp$ have been repeated with $A_\perp = 0$ and little difference was observed, confirming that in our experiments $A_\perp$ is not substantially deteriorating the effectiveness of the spectator qubit approach.

\section{SUPPLEMENTARY NOTE: Spectator readout basis and phase compensation}
\label{sec:SI_Bayesian}
According to Bayes' theorem, the posterior probability density distribution $P(H|E)$ of a parameter $H$ given that evidence $m$ has been observed, is given by 

\begin{equation}
    P(H|m) = \frac{ P(m|H) \cdot P(H) }{P(m)},
\end{equation}

\noindent where $P(m|H)$ is the probability to obtain evidence $m$ given a parameter $H$, $P(H)$ is the prior probability density function and $P(m) = \int P(m|H) P(H) dH$. \\

For a quantum state given by $\ket{\psi} = \frac{1}{\sqrt{2}} \left( \ket{0} + \exp(i g \phi) \ket{1}\right)$, a measurement in a basis $\theta$ in the XY-plane obtains an outcome $m = {0,1}$ with probability $\frac{1}{2} \left(1 + (-1)^m \cos{\left( g \phi - \theta \right)}\right)$. By induction, for $n$ spectators, this gives equation \eqref{eq:Bayesian}.

\section{SUPPLEMENTARY NOTE: Spectator readout basis and phase compensation}
\label{sec:RObasis_phi_c}

\noindent Dephasing is optimally mitigated if the uncertainty of the memory qubit phase probability density function (PDF) $P(\phi)$ is minimized. The uncertainty over a cyclic variable is described by the Holevo phase variance \cite{terhal_encoding_2016,berry_optimal_2001}:

\begin{equation}
    V[P(\phi)] = S[P(\phi)]^{-2}-1 \equiv \left| \int e^{i\phi} P(\phi) d\phi \right|^{-2} - 1,
    \label{eq:Holevo}
\end{equation}

\noindent with $S[P(\phi)]\in[0,1]$ the \textit{sharpness} parameter. One can show that for a small phase variance $V[P(\phi)] \approx \langle (\phi - \langle \phi \rangle)^2\rangle$, being the usual variance for a non-cyclic parameter \cite{berry_optimal_2001}. The optimal readout phase $\theta_{m}$ of the m'th spectator is the one that, given a prior distribution defined by previous measurement outcomes $x[m-1]$, maximizes the sharpness parameter of the posterior PDF averaged over the possible readout outcomes $x_m$: 

\begin{equation}
    \theta_{m}^{\mathrm{opt}} = \underset{\theta_m}{\arg \max} \, \left\{ S_{posterior}(\theta_m) \right\} = \underset{\theta_m}{\arg \max} \left\{ \sum_{x_m=0,1} P_{\theta_m}(x_m|x[m-1]) \cdot S[P_{\theta_m}(\phi|x[m])] \right\}.
    \label{eq:phi_m_optimal_1}
\end{equation}

\noindent Here $P_{\theta_m}(x_m|x[m-1])$ is the probability of obtaining measurement outcome $x_m$ on the m'th measurement in a readout basis defined by $\theta_m$ given a measurement outcome $x[m-1]$ of previous measurements in bases defined by $\{\theta_1,...,\theta_{m-1}\}$, which we assume to be fixed. $P_{\theta_m}(\phi|x[m])$ is the PDF of $\phi$ conditioned on measurement outcomes $x[m]$. Using Bayes rule and the conditional probability identity $P(C|AB) = P(CB|A)/P(B|A)$ we write \cite{terhal_encoding_2016}

\begin{align}
    \begin{split}
        P_{\theta_m}(\phi|x[m]) & = \frac{P(\phi|x[m-1]) \cdot P_{\theta_m}(x_m|\phi,x[m-1])}{P_{\theta_m}(x_m|x[m-1])} \\
                            & = \frac{P(\phi|x[m-1]) \cdot P_{\theta_m}(x[m]|\phi)}{P_{\theta_m}(x_m|x[m-1]) \cdot P(x[m-1]|\phi)}. \\
        \label{eq:posterior_phi}
    \end{split}
\end{align}

\noindent Inserting equation \eqref{eq:posterior_phi} into equation \eqref{eq:phi_m_optimal_1} leads to:

\begin{equation}
    \theta_{m}^{\mathrm{opt}} = \underset{\theta_m}{\arg \max} \sum_{x_m=0,1} \left| \int e^{i\phi} \frac{P(\phi|x[m-1]) \cdot P_{\theta_m}(x[m]|\phi)}{P(x[m-1]|\phi)} d\phi \right|.
    \label{eq:phi_m_optimal_2}
\end{equation}

\noindent Given a set of measurement bases $\theta=\{\theta_i\}$ and a state with angle $\phi$, the measurement results of each measurement are independent and hence

\begin{equation}
    P_{\theta_m}(x[m]|\phi) = \prod_{i=1}^m P_{\theta_i}(x_i|\phi) = \prod_{i=1}^m \cos^2\left(\frac{\phi-\theta_i}{2} + x_i\frac{\pi}{2}\right),
    \label{eq:P_x[m]_given_phi}
\end{equation}
where the measurement result $x_i$ is either 0 or 1. Similarly equation \eqref{eq:P_x[m]_given_phi} holds up for $P(x[m-1]|\phi)$ and hence equation \eqref{eq:phi_m_optimal_2} reduces to

\begin{align}
    \begin{split}
        \theta_{m}^{\mathrm{opt}} &= \underset{\theta_m}{\arg \max} \,\left\{ S_{posterior}(\theta_m) \right\} \\
        &= \underset{\theta_m}{\arg \max} \left\{ \sum_{x_m=0,1} \left| \int e^{i\phi} \cdot P(\phi|x[m-1]) \cdot \cos^2\left(\frac{\phi-\theta_m}{2} + x_m\frac{\pi}{2}\right) d\phi \right| \right\}.
        \label{eq:phi_m_optimal_3}
    \end{split}
\end{align}

\begin{figure*}[!ht]
\begin{centering}
\includegraphics[width=\textwidth]{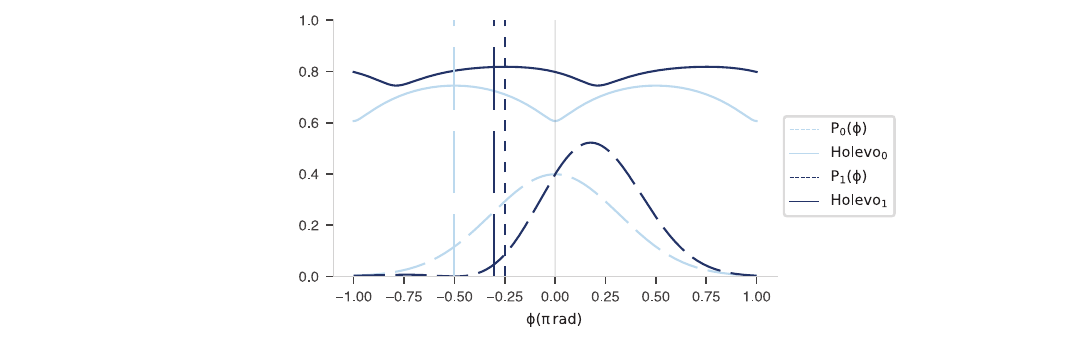}
\caption{\label{fig:SI_fig_OptimalROangle} \textbf{Optimal spectator readout angle.} Dashed curves correspond to phase PDFs $P(\phi)$ and the solid lines correspond to $S_{posterior}(\phi)$. Large-dashed vertical lines indicate $\langle \overline{\phi} \rangle + \frac{\phi}{2}$ (see below equation \eqref{eq:phi_m_optimal_3}). The small-dashed vertical line indicates $\phi_m^{\mathrm{opt}}$. The underlying PDFs are $P(\phi) = \frac{1}{\sqrt{2\pi}\sigma} e^{-\frac{\phi^2}{2\sigma^2}} $ (light blue) and the posterior PDF after a measurement in a basis $\langle \overline{\phi} \rangle + \frac{\phi}{2}$ (see equation \eqref{eq:Bayesian}, given by $P(\phi) = \frac{1}{\sqrt{2\pi}\sigma} e^{-\frac{\phi^2}{2\sigma^2}} \cdot \left( 1 + \cos{\left( \phi-\frac{\pi}{2} \right)} \right)$ (dark blue). In the data shown here $\sigma=1$. For the light blue data, the dashed and dotted line overlap. For the dark blue data, the readout perpendicular to $\langle \overline{\phi} \rangle$ is sub-optimal, but the decrease in the sharpness parameter $S_{posterior}(\phi)$ is minor.}
\end{centering}
\end{figure*}

\noindent If $P(\phi|x[m-1])$ is a normal distribution with a expectation value $ \langle \overline{\phi} \rangle = \arctan{ \left(\frac{\langle \sin{\phi} \rangle}{\langle \cos{\phi} \rangle} \right)}  = \arctan{ \left( \frac{\langle Y \rangle}{\langle X \rangle} \right)}$, then $\theta_{m}^{\mathrm{opt}} = \langle \overline{\phi} \rangle \pm \frac{\pi}{2}$. The stochastic electron-spin evolution during the entanglement attempts imparts a random-walk-like correlated evolution on the nuclear spins (SI section \ref{sec:SI_1Spec_analytical}). For large numbers of entanglement attempts, the central limit theorem provides a normally distributed nuclear phase probability density function. For the first spectator qubit, this justifies the choice of a readout basis perpendicular to the Bloch vector angle. We also use this strategy for spectator $M>1$, which according to equation \eqref{eq:phi_m_optimal_3} is not necessarily optimal. However, for the second spectator, the difference in sharpness parameter between the most optimal readout basis and a readout bases set by $\langle \overline{\phi} \rangle + \frac{\pi}{2}$ is minor. This is illustrated in SI figure \ref{fig:SI_fig_OptimalROangle} where the initial PDF is a normal distribution with $\sigma=1$.\\

Having obtained a measurement result, the real-time feedback allows to apply a phase rotation $\phi_c$ on the memory qubit and on the other spectator qubits. We want to choose $\phi_c$ to optimally recover the memory qubit state, which we express by its fidelity with respect to the X-axis ($\theta=0$). A single spectator measurement gives two measurement outcome-dependent PDFs on which an independent phase rotation can be applied. Hence the fidelity we target to maximize is described as

\begin{align}
    \begin{split}
        F(\{\phi_{c_i}\}) = &p_{\theta_m}(0) \int_{-\infty}^{\infty} \frac{1}{2} \left[ 1 + \cos{(\phi)} \right] \cdot P_{\theta_m}(\phi-\phi_{c_0}|0)d\phi \\
        &+ p_{\theta_m}(1) \int_{-\infty}^{\infty} \frac{1}{2} \left[ 1 + \cos{(\phi)} \right] \cdot P_{\theta_m}(\phi-\phi_{c_1}|1)d\phi, 
        \label{eq:Fidelity_phi_c_raw}
    \end{split}
\end{align}

\noindent where $P_{\theta_m}(\phi|i)$ is the posterior PDF of $P(\phi)$ after performing a measurement along basis $\theta_m$ and obtaining measurement outcome \textquotedbl$i$\textquotedbl. $p_{\phi_m}(i)$ is the probability to obtain measurement outcome \textquotedbl$i$\textquotedbl\, and $\phi_{c_i}$ is the corresponding phase rotation. Using Bayes' theorem (see section \ref{sec:SI_Bayesian}) we can write

\begin{equation}
    P_{\theta_m}(\phi|i) = \frac{\frac{1}{2}\left[1+\cos{(\phi-\theta_m + i\pi)}\right] \cdot P(\phi)}{P_{\theta_m}(i)},
    \label{eq:Bayesian_SI}
\end{equation}

\noindent and hence for a chosen $\theta_m$ the most optimal set $\{\phi_{c_i}^{\mathrm{opt}}\}$ of phase compensation angles is given by

\begin{align}
    \begin{split}
    \phi_{c_i}^{\mathrm{opt}} &= \underset{\phi_{c_i}}{\arg \max} \left\{ F(\{\phi_{c_i}\}) \right\} \\&= \underset{\phi_{c_i}}{\arg \max} \left\{ \sum_i \int_{-\infty}^{\infty} \frac{1}{2}\left[1+\cos{(\phi)}\right] \cdot \frac{1}{2}\left[1+\cos{(\phi - \phi_{c_i} - \theta_m +i\pi)}\right] \cdot P(\phi - \phi_{c_i}) \, d\phi \right\}.
    \label{eq:Fidelity_phi_c_optimal}
    \end{split}
\end{align}

\noindent For a symmetric phase PDF $P(\phi)$ it holds that $\phi_{c_1} = -\phi_{c_0}$. Considering a normal phase PDF $P(\phi)$ and an optimal readout angle $\theta_m = \frac{\pi}{2}$, equation \eqref{eq:Fidelity_phi_c_optimal} gives $\phi_{c_0} = \arctan{\left(-e^{-\frac{\sigma^2}{2}}\sinh{\left(\sigma^2\right)}\right)}$, with $\sigma$ the standard deviation of the normal distribution.

\section{SUPPLEMENTARY NOTE: Single spectator effect: analytical expressions}
\label{sec:SI_1Spec_analytical}

In this section we will derive an analytical expression for the memory qubit fidelity upon using a single spectator after running $\mathrm{N_{REA}}$ entanglement sequences as given in figure \ref{Figure2}a. We will assume that both the memory and spectator qubit have zero $A_\perp$ and their parallel hyperfine components are related by $g = \frac{A_{\parallel,s}}{A_{\parallel,m}}$. Additionally, we assume perfect gates and instantaneous electron reinitialization. Correspondingly, the only source of dephasing is given by the projection of the electron-spin qubit to the $\ket{0}$ or $\ket{1}$ state after each entanglement attempt with respective probabilities $p_0$ and $p_1 = 1-p_0$. Note that other dephasing mechanisms, like a stochastic reset time for the electron spins would impart dephasing following a similar result.  Here we will use $p_0 = \frac{1}{2}$, which corresponds to a $\frac{\pi}{2}$ pulse to generate an electron-spin superposition state at the start of the entanglement sequence. \\

Generally, the fidelity of a quantum state in the XY-plane with a phase probability density $P(\phi)$ is given by

\begin{equation}
    F_\theta = \int_{-\infty}^{\infty} \frac{1}{2} \left[1+\cos\left(\phi - \theta\right) \right] P(\phi)d\phi,
    \label{eq:Spectator_fidelity_0}
\end{equation}

\noindent where $\theta$ defines the angle in the XY plane along which the fidelity is measured. In the following, we target to determine $P(\phi)$ and $\theta_c$, the optimal angle along which to measure the fidelity. We will start with with $P(\phi)$ and consider each nuclear-spin qubit in their respective rotating frames with the respective frequency being the average of the two electron-spin-dependent frequencies ($f = \frac{1}{2}(f_{m_s=0} + f_{m_s=1})$). Following the assumptions above, the nuclear-spin qubits pick up a phase $\pm \frac{1}{2} A_\parallel t_e$ at each entanglement attempt, such that after N (here N=$\mathrm{N_{REA}}$) entanglement attempts the phase distribution is given by:

\begin{equation}
    P(\phi,N) = \sum_{M=0}^N \frac{1}{2^M} \frac{N!}{M!(N-M)!} \delta\left(\phi - M A_\parallel t_e \right).
\end{equation}

\noindent For large N, this binomial distribution is expressed as 

\begin{equation}
    P(\phi,N) = \sqrt{\frac{2}{\pi N A_\parallel^2 t_e^2}} \exp\left(-2 \frac{\phi^2}{N A_\parallel^2 t_e^2}\right) = \frac{1}{\sigma \sqrt{2 \pi}} \exp\left( -\frac{\phi^2}{2\sigma^2} \right), \\ 
\end{equation}

\noindent with $\sigma = \sqrt{N}A_\parallel t_e$.\\

After measuring a spectator with a g-value equal to "g", the posterior phase distribution is given by (see equation \eqref{eq:Bayesian}) $P(\phi|H) = \left[1+(-1)^H \cos(g\phi - \chi) \right] \cdot P(\phi,N)$, with H the spectator-measurement outcome and $\chi$ the spectator readout phase, for which we use $\chi = \langle g \phi \rangle +\frac{\pi}{2}$ (see section \ref{sec:RObasis_phi_c}) calculated using the phase distribution prior to a spectator readout. The spectator measurement both narrows and shifts the memory phase distribution, such that the expected spectator-dependent memory-qubit phase ($\langle \phi_{H} \rangle $) is non-zero. Experimentally, we target to measuring the memory qubit fidelity along $\theta_c = \langle \phi_{H} \rangle$, where 

\begin{subequations}
    \begin{align}
        \langle \phi_{H} \rangle &= \arctan\left( \frac{\langle \sin(\phi) \rangle}{\langle \cos(\phi) \rangle} \right)
        \label{eq:1spec_phi_BV}\\
        \langle \sin(\phi) \rangle &= \int_{-\infty}^{\infty} \sin(\phi)\left[ 1 + (-1)^H \sin(g \phi) \right] \frac{1}{\sigma \sqrt{2 \pi}} \exp\left( -\frac{\phi^2}{2\sigma^2} \right) \notag \\ 
        &= (-1)^H \cdot \frac{1}{2} \left[ \exp\left(-\frac{1}{2}(g-1)^2\sigma^2\right) - \exp\left(-\frac{1}{2}(g+1)^2\sigma^2\right) \right]
        \label{eq:1spec_sin_phi}\\
        \langle \cos(\phi) \rangle &= \int_{-\infty}^{\infty} \cos(\phi)\left[ 1 + (-1)^H \sin(g \phi) \right] \frac{1}{\sigma \sqrt{2 \pi}} \exp\left( -\frac{\phi^2}{2\sigma^2} \right) \notag\\
        &= \exp\left( -\frac{\sigma^2}{2} \right).
        \label{eq:1spec_cos_phi}
    \end{align}
\end{subequations}

\noindent Inserting equation \eqref{eq:1spec_sin_phi} and \eqref{eq:1spec_cos_phi} in equation \eqref{eq:1spec_phi_BV} gives

\begin{equation}
    \langle \phi_{H} \rangle = (-1)^H \arctan\left( \frac{1}{2} \exp\left( -\frac{1}{2} g \left(g+2\right)\sigma^2 \right) \cdot \left(\exp\left(2 g \sigma^2\right)-1 \right) \right),
    \label{eq:phi_H}
\end{equation}

\noindent which combined with equation \eqref{eq:phi_H} with equation \eqref{eq:Spectator_fidelity_0} and equation \eqref{eq:Bayesian} with the spectator readout axis equal to $\pi/2$ (orthogonal to the predicted Bloch vector) gives the fidelity averaged over both spectator-measurement outcomes as

\begin{equation}
    F_\theta = \sum_H \frac{1}{2} \int_{-\infty}^{\infty} \frac{1}{2} \left[1+\cos\left(\phi - \langle \phi_{H} \rangle \right) \right] \cdot \left[1+(-1)^H \sin(g\phi) \right] \frac{1}{\sigma \sqrt{2 \pi}} \exp\left( -\frac{\phi^2}{2\sigma^2} \right) d\phi. \\
    \label{eq:Spectator_fidelity_1}
\end{equation}

\noindent Evaluating this expressing allows to rewrite $F_\theta$ in a simpler form given by 

\begin{subequations}
    \begin{align}
        F_\theta &= \frac{1}{2} + \frac{1}{4}\sqrt{\rho} \exp\left(-\frac{1}{2}\sigma^2 \right)
        \label{eq:Spectator_fidelity_2}\\
        \rho &= 4 + \frac{(\nu - 1)^2}{\mu \nu}\\
        \nu &= \exp\left(2 g \sigma^2 \right)\\
        \mu &= \exp\left(g^2 \sigma^2\right),
    \end{align}
\end{subequations}

\noindent which reduces to $F_\theta = \frac{1}{2} + \frac{1}{2}\exp\left( -\frac{1}{2} \sigma^2 \right)$ for $g=0$, being equal to the spectator-less situation. \\

\begin{figure*}[!ht]
\includegraphics[width=\textwidth]{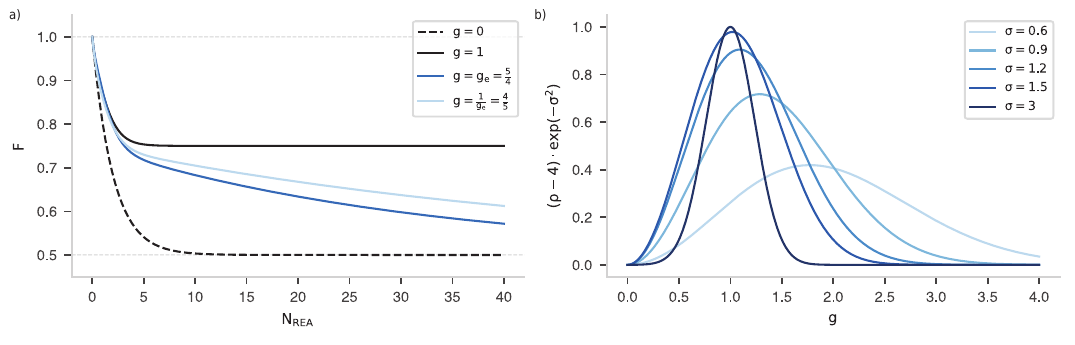}
\caption{\label{fig:SI_fig_1Spec_analytical} \textbf{Single spectator effect.} \textbf{a)} Fidelity of the memory qubit after running $\mathrm{N_{REA}}$ entanglement attempts and using one spectator with different g-values. The gate sequence of each attempt is shown in figure \ref{Figure2}a. The fidelity is the fidelity measured in the XY plane with the memory qubit initialized in the XY plane. \textbf{b)} The term in the fidelity responsible for the improvement over the case without a spectator. For memory distributions with a small dephasing, larger g-values are desirable, whereas for large dephasing, g-values equal to 1 become preferential. }
\end{figure*}

In SI figure \ref{fig:SI_fig_1Spec_analytical}a, we plot the equation \eqref{eq:Spectator_fidelity_2} for different g-values and $\sigma=1$. We observe that for g=1, so a spectator qubit identical to the memory qubit, the fidelity can be restored from $\frac{1}{2}$ to $\frac{3}{4}$. In light and dark blue, we plot the fidelity for $g=g_e$ and $g=\frac{1}{g_e}$ from which a larger improvement is observed for $g=\frac{1}{g_e}$. We hypothesise that the quick initial decay is related to the memory qubit $A_\parallel$, while the following slower decay is related to timescale needed for the difference between the parallel hyperfine components of the memory and spectator qubit to acquire a phase on the order of 2$\pi$ which prohibits unique identification of the memory phase upon measuring the spectator qubit phase. \\

In SI figure \ref{fig:SI_fig_1Spec_analytical}b, $(\rho-4)\exp\left(-\sigma^2\right)$ is plotted. This effectively describes the possible improvement possible in $F_\theta$ as a function of g, plotted for different values of $\sigma$. The plot shows that for low $\sigma$ one benefits from a larger g-value, as then the spectator is susceptible to the small phases picked up by the memory qubit. However, for larger $\sigma$, the optimal g-value approaches g=1, to avoid a 2$\pi$ phase wrapping following from the differences of the memory and spectator $A_\parallel$.

\subsection{Fitting experimental data}
\label{sec:SI_fitparameters}
To fit the data measured in this work where the memory qubit was initialized in the XY plane, we modify the function of equation \eqref{eq:Spectator_fidelity_2} to include SPAM errors. As the spectator approach is designed to compensate correlated dephasing, we use an exponential decay as a fit function for data with the memory qubit initialized in the Z-basis. Correspondingly, for these two cases, the fit functions are given by

\begin{subequations}
    \begin{align}
        F_{xy} &= \frac{1}{2} + \frac{1}{4} (1-\mathrm{A_{SPAM}}) \sqrt{\rho} \exp\left(-\frac{1}{2}\sigma^2 \right)\\
        \label{eq:Spectator_fidelity_3_fit}
        F_{z} &= (1-\mathrm{A_{SPAM_z}}) \exp\left(- \frac{ \mathrm{N_{REA}} }{N_{1/e}} \right) + a,
    \end{align}
\end{subequations}

\noindent where we substitute $\sigma = \sqrt{ \mathrm{N_{REA}} A_\parallel t_e}$. We consider $A_\parallel t_e$ as a single fit parameter. For the fits of $F_{xy}$, we fix the g-value to 0 if no spectator is used, to the g-value of the spectator if a single spectator is used and leave it as a free parameter if more than one spectator are used. For $F_{z}$, we fix $a=\frac{1}{2}$. \\

The fit parameters of the measurement-based data in figure \ref{Figure2}d shown in table \ref{tab:SI_msmt_based_fitparams}:

\begin{table}[ht]
\centering
\begin{adjustbox}{max width=\textwidth}
\begin{tabular}{|c|c|c|c|c|c|c|c|c|c|}
\hline
\textbf{K spectators} & \textbf{$\mathrm{A_{SPAM}}$} & \textbf{g} & \textbf{$\mathrm{A_\parallel t_e}$} & \textbf{$\chi^2_{xy}$} \\ \hline
K = 0 & 0.082 $\pm$ 0.007 & fixed, g = 0 & 0.0271 $\pm$ 0.0003 & 5.06 \\ 
K = 1 & 0.168 $\pm$ 0.007 & fixed, g = 1.49 & 0.0309 $\pm$ 0.0004 & 2.18 \\ 
K = 2 & 0.279 $\pm$ 0.008 & 0.68 $\pm$ 0.02 & 0.025 $\pm$ 0.001 & 2.91 \\ \hline
\textbf{K spectators} & \textbf{$\mathrm{A_{SPAM_z}}$} & \textbf{a} & \textbf{$\mathrm{N_{1/e}}$} & \textbf{$\chi^2_{z}$} \\ \hline
K = 0 & 0.105 $\pm$ 0.008 & fixed, a = 0.5 & 2.1e3 $\pm$ 0.3e3 & 1.66 \\ 
K = 1 & 0.122 $\pm$ 0.005 & fixed, a = 0.5 & 4.4e4 $\pm$ 0.7e3 & 3.60 \\ 
K = 2 & 0.110 $\pm$ 0.005 & fixed, a = 0.5 & 3.1e3 $\pm$ 0.3e3 & 2.13 \\ 
\hline
\end{tabular}
\end{adjustbox}
\caption{\label{tab:SI_msmt_based_fitparams} fit parameters of fits through data in figure \ref{Figure2}c.}
\end{table}

\noindent The fit parameters of the gate-based data in figure \ref{Figure4}a shown in table \ref{tab:SI_gate_based_fitparams}:

\begin{table}[ht]
\centering
\begin{adjustbox}{max width=\textwidth}
\begin{tabular}{|c|c|c|c|c|c|c|c|c|c|}
\hline
\textbf{K spectators} & \textbf{$\mathrm{A_{SPAM}}$} & \textbf{g} & \textbf{$\mathrm{A_\parallel t_e}$} & \textbf{$\chi^2_{xy}$} \\ \hline
K = 0 & fixed, 0.066 & fixed, g = 0 & 0.0271 $\pm$ 0.0002 & 9.53 \\ 
K = 1 & fixed, 0.081 & fixed, g = 1.49 & 0.0314 $\pm$ 0.0002 & 5.01 \\ 
K = 2 & fixed, 0.125 & 0.588 $\pm$ 0.009 & 0.0300 $\pm$ 0.0004 & 8.15 \\ \hline
\textbf{K spectators} & \textbf{$\mathrm{A_{SPAM_z}}$} & \textbf{a} & \textbf{$\mathrm{N_{1/e}}$} & \textbf{$\chi^2_{z}$} \\ \hline
K = 0 & fixed, 0.1033 & fixed, a = 0.5 & 1.95e3 $\pm$ 0.09e3 & 34.62 \\ 
K = 1 & fixed, 0.1141 & fixed, a = 0.5 & 3.0e3$ \pm$ 0.1e3 & 10.05 \\ 
K = 2 & fixed, 0.0961 & fixed, a = 0.5 & 2.6e3 $\pm$ 0.1e3 & 3.04 \\ 
\hline
\end{tabular}
\end{adjustbox}
\caption{\label{tab:SI_gate_based_fitparams} fit parameters of fits through data in figure \ref{Figure4}a.}
\end{table}

To fit the data of Fig.~\ref{Figure2}b, we multiply equation \eqref{eq:1spec_sin_phi} (for $\langle Y \rangle$) and equation \eqref{eq:1spec_cos_phi} (for $\langle X \rangle$) with $(1-A_{SPAM})$ and use the same substitution of $\sigma$ as above. We obtain the following fitparameters, where the difference in $A_{SPAM}$ is explained by the asymmetry in electron-spin readout fidelity.

\begin{table}[ht]
\centering
\begin{adjustbox}{max width=\textwidth}
\begin{tabular}{|c|c|c|c|c|c|c|c|c|c|}
\hline
\textbf{$\langle X \rangle$} & \textbf{$\mathrm{A_{SPAM}}$} & \textbf{g} & \textbf{$\mathrm{A_\parallel t_e}$} & \textbf{$\chi^2_{xy}$} \\ \hline
black & 0.21 $\pm$ 0.01 & fixed, g = 1.49 & 0.0276 $\pm$ 0.0005 & 1.62 \\ \hline 
\textbf{$\langle Y \rangle$} & \textbf{$\mathrm{A_{SPAM}}$} & \textbf{g} & \textbf{$\mathrm{A_\parallel t_e}$} & \textbf{$\chi^2_{xy}$} \\ \hline
green & 0.02 $\pm$ 0.03 & fixed, g = 1.49 & 0.033 $\pm$ 0.001 & 3.63 \\ 
red & 0.26 $\pm$ 0.03 & fixed, g = 1.49 & 0.034 $\pm$ 0.002 & 2.49 \\ 
\hline
\end{tabular}
\end{adjustbox}
\caption{\label{tab:SI_X_Y_expval_fitparams} fit parameters of fits through data in figure \ref{Figure2}b. The color in the table refers to for which data and line the fitparameters correspond.}
\end{table}

To fit the data of Fig.~\ref{Figure3}d, we divide equation \eqref{eq:phi_H} by $2\pi A_{\parallel,m}$ to obtain

\begin{equation}
    \delta_{t,opt} = \frac{1}{2\pi A_{\parallel,m}} \cdot 2 \cdot \arctan\left( \frac{1}{2} \exp\left( -\frac{1}{2} g \left(g+2\right)\sigma^2 \right) \cdot \left(\exp\left(2 g \sigma^2\right)-1 \right) \right),
\end{equation}
where the factor 2 in front of the $\arctan$ is because the phase difference between the red and green distributions in Fig.~\ref{Figure3}b is the difference between the two different readout outcomes ($H={0,1}$) of equation \eqref{eq:phi_H}. Again, we substitute $\sigma = \sqrt{\mathrm{N_{REA}} A_\parallel t_e}$, fix $g=1.49$ as the ratio between the parallel hyperfine parameters of spectator and memory qubit and fix $A_{\parallel_m} = 24.4$ kHz. From the fit, we obtain $A_\parallel t_e$ = 0.0280(6) and $\chi^2$ = 4.70.

\section{SUPPLEMENTARY NOTE: Measurement-induced dephasing}
\label{sec:Measurement_Dephasing}
\begin{figure*}[!ht]
\begin{centering}
\includegraphics[width=\textwidth]{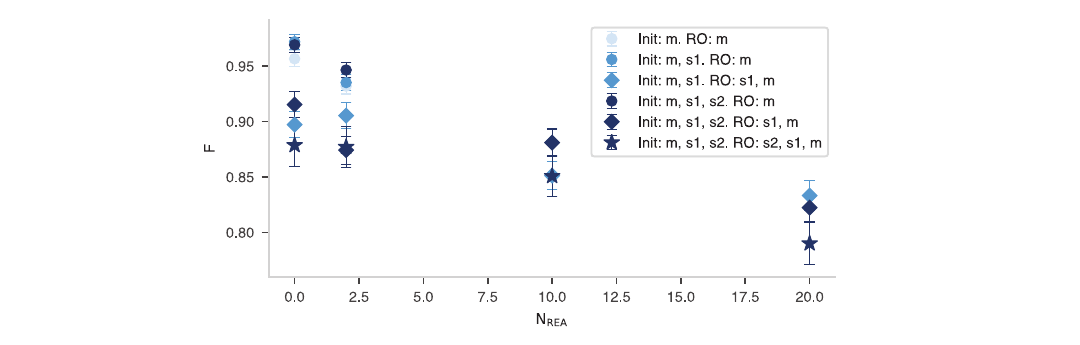}
\caption{\label{Fig:SI_readout_dephasing} \textbf{Readout induced dephasing.} Fidelity with the +X basis of the memory qubit initialized in the $\mathrm{\mathrm{\ket{+X}}}$ state after $\mathrm{N_{REA}}$ entanglement generation attempts for different spectator initialization and readout strategies. The data is part of the dataset in figure \ref{Figure2}d. The coding in legend uses `m' for memory qubit, `sx' for spectator x, `Init' to indicate which qubits are initialised and `RO' to indicate which qubits are read out. Initialising additional spectator qubits barely affects the memory qubit fidelity at small N, while the readout of these spectators does decrease the fidelity.}
\end{centering}
\end{figure*}

The experimental dataset shown in figure \ref{Figure2}c only shows data where M = $\{0,1,2\}$ spectator qubits are initialised and read out. The BVL at $\mathrm{N_{REA}}$=0 decreases for an increasing amount of spectators used. Thus, the initialization and read out of the spectators provide an additional dephasing mechanism. Here, alongside the data shown in figure \ref{Figure2}c, we show a complementary dataset where we initialise M spectators and read out K$\leq$M spectators. The data is shown in SI figure \ref{Fig:SI_readout_dephasing}. C0 is the memory qubit, C1 spectator qubit 1 and C2 spectator qubit 2 (see figure \ref{Figure2}a). Points indicated with circles represent data where only the memory qubit is read out. Points represented by diamonds represent data where only spectator 1 and the memory qubit are read out, and stars represent data where subsequently spectator 2, spectator 1 and the memory qubit are read out. For the light blue data only the memory qubit is initialised, while for the blue data also spectator 1 is initialised and for the dark blue data both spectators are initialised. The data shows that initialising more spectators does not significantly reduce the memory qubit BVL, but reading out more spectators does reduce the memory qubit BVL. 

\FloatBarrier

\section{SUPPLEMENTARY NOTE: Simulation background}
\label{sec:Simulation_Derivation}

\begin{figure*}[!ht]
\includegraphics[width=\textwidth]{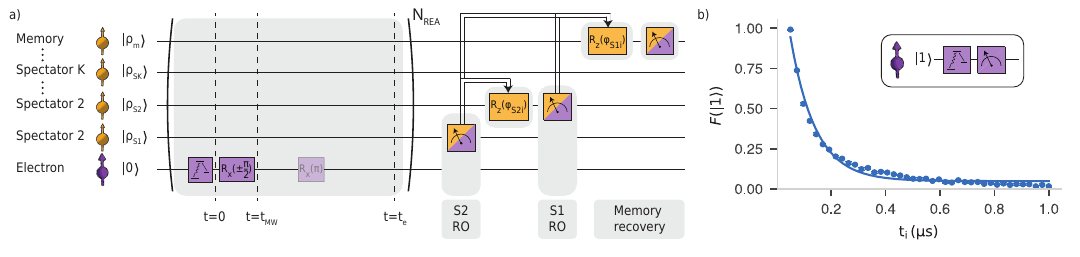}
\caption{\label{fig:SI_fig_Entanglement_Sequence_Simulation} \textbf{General entanglement sequence and spectator qubit implementation.} \textbf{a)} Upon the start of an entanglement sequence, the electron is initialized in $\ket{0}$ and subsequently rotated to an (unbalanced) superposition state to allow spin-photon entanglement (we omit the optical pi-pulse here). Subsequent coherent rotations on the electron can be performed, as indicated by the transparent $\pi$-pulse. Upon completion of $\mathrm{N_{REA}}$ entanglement attempts, a set of spectators is read out and feedforward is applied. \textbf{b)} Fidelity with respect to the $\ket{1}$ state of the electron-spin qubit upon applying a repump laser for time $t_i$ after initializing the electron in $\ket{1}$. The dataset shown here uses a repump power of $\SI{600}{\nano\watt}$, alike all measurements presented in the main text. We fit a single exponential decay $F(\ket{1}) = A\, \exp{\left(- t/t_i\right)} + a$ and extract $t_i = 92 \pm 4 \SI{}{\nano\second}$.}
\end{figure*}

We adopt a density matrix formalism approach to simulate the spectator qubit approach. In general, an entanglement sequence starts with an electron spin initialization to $\ket{0}$ followed by an (unbalanced) microwave $\pi$ pulse that prepares the electron spin in a superposition required for spin-photon entanglement. Following this MW pulse, an optional electron decoupling is applied (not done the experiments presented). Finally a set of spectator qubits is read out based on which real-time feedback is applied on other spectator qubits and the memory qubit, see SI figure \ref{fig:SI_fig_Entanglement_Sequence_Simulation}. \\

Upon the start of the sequence the total density matrix equals $\rho_{t=0,N=0} =  \prod_i \rho_{Si} \otimes \rho_m \otimes \ket{0}_e\bra{0}_e$, where the product refers to a tensor product of different Hilbert spaces, $\rho_{Si}$ to the density matrix of spectator number $i$, $\rho_m$ to the memory qubit density matrix and the subscript $e$ to the electron state. We define the start of the entanglement sequence ($t=0$) after the electron reinitialization (see SI figure \ref{fig:SI_fig_Entanglement_Sequence_Simulation}). The subsequent MW pulse along the X-axis is modelled as an instantaneous pulse and creates an unbalanced superposition defined by $\alpha$ on the electron-spin qubit. The MW pulse modifies the density matrix to  

\begin{align}
    \begin{split}
        \rho(t = t_{MW}) = &\rho_{t=0,N=0} \, \otimes \\
        &\left[ \abs{\alpha}^2\ket{0}_e\bra{0}_e - i\alpha\sqrt{1-\abs{\alpha}^2}\ket{0}_e\bra{1}_e + i\alpha^\dagger\sqrt{1-\abs{\alpha}^2}\ket{1}_e\bra{0}_e + (1-\abs{\alpha}^2) \ket{1}_e\bra{1}_e \right].
        \label{eq:SI_rho_post_MW}
    \end{split}
\end{align}

\noindent After the MW pulse, there is a free evolution of time $t_e$ followed by the application of a laser pulse of time $t_i$ that targets to reinitialise the electron to $\ket{0}$. This reinitialization is stochastic and at each $\mathrm{N_{REA}}$ we sample the time $\tau_\mathrm{N}$ at which the $\ket{1}$ state is reinitialised to $\ket{0}$ from a single exponential decay with decay constant $\tau_d$, see SI figure \ref{fig:SI_fig_Entanglement_Sequence_Simulation}b. While the decay is more accurately described by a sum of two exponential decays (see SI figure \ref{fig:SI_fig_Repump_Decay}), the single exponential decay can be easily inverted to sample $\tau$. The density matrix in our model at time $T = t_e + \tau_\mathrm{N}$ after the MW-pulse is hence described by\\

\begin{subequations}
    \begin{align}
        U_{evo,T} =& U_{0,T} \otimes \ket{0}_e\bra{0}_e + U_{1,T} \otimes \ket{1}_e\bra{1}_e \label{eq:SI_LDE_evo}\\
        \begin{split}
        \rho_{t=(t_e+\tau_\mathrm{N})} =& \abs{\alpha}^2 U_{0,T} \rho_{t=0,N=0} U_{0,T}^\dagger \ket{0}_e\bra{0}_e - i\alpha\sqrt{1-\abs{\alpha}^2} U_{0,T} \rho_{t=0,N=0} U_{1,T}^\dagger \ket{0}_e\bra{1}_e + \\
        &i\alpha^\dagger\sqrt{1-\abs{\alpha}^2} U_{1,T} \rho_{t=0,N=0} U_{0,T}^\dagger \ket{1}_e \bra{0}_e + (1-\abs{\alpha}^2) U_{1,T} \rho_{t=0,N=0} U_{1,T}^\dagger \ket{1}_e\bra{1}_e,
        \label{eq:SI_LDE_rho}
        \end{split}
    \end{align}
\end{subequations}

\noindent where $U_i$ only acts on the nuclear spins Hilbert space and is determined by the electron spin state during the entanglement attempt and equation \eqref{eq:Hamiltonian_n_electronDependent}. Subsequently, we model the electron reinitilisation by tracing out the electron state in equation \eqref{eq:SI_LDE_rho} and a free evolution time of $T_0 = t_i - \tau_N$ with the electron in the $\ket{0}$ state. Therefore, after one entanglement generation sequence ($\mathrm{N_{REA}}$=1), we have

\begin{equation}
    \rho_{t=0,\mathrm{N_{REA}}=1} = U_{0,T_0}\left[ \abs{\alpha}^2 U_{0,T} \rho_{t=0,N=0} U_{0,T}^\dagger + (1-\abs{\alpha}^2) U_{1,T} \rho_{t=0,N=0} U_{1,T}^\dagger \right] U_{0,T_0}^\dagger \ket{0}_e\bra{0}_e.
    \label{eq:SI_rho_1_entanglement_round}
\end{equation}

\noindent The final density matrix $\rho_{\mathrm{N_{REA}}}$ after $\mathrm{N_{REA}}$ repetitions of the entanglement sequence is given by iterating through the protocol from equation \eqref{eq:SI_rho_post_MW} to \eqref{eq:SI_rho_1_entanglement_round} $\mathrm{N_{REA}}$ times. This can be straightforwardly done by substituting $\rho_{t=0,\mathrm{N_{REA}}=0}$ in equation \eqref{eq:SI_rho_post_MW} for the output of equation \eqref{eq:SI_rho_1_entanglement_round}.\\

After having obtained $\rho_f$, we proceed by implementing the spectator protocol. The target is to read out spectator $k$ in a basis perpendicular to the measurable expectation value $ \langle \overline{\phi} \rangle_k$ (see SI section \ref{sec:RObasis_phi_c}). We calculate $\langle \overline{\phi} \rangle_k$ from $\rho_N$, apply a $R_z(\phi-\langle \overline{\phi} \rangle_k)$ rotation on spectator $k$ and subsequently read out spectator $k$ in the Y-basis. In case of perfect readout, the probability ($P_{k_i}$) and corresponding density matrices ($\rho_{k_i}$) conditioned on obtaining a bright/dark measurement outcome on spectator $k$ are given by

\begin{subequations}
    \begin{equation}
      \rho_{k_b} = P_{k_b}^{-1} \left( \rho_{k_y} \otimes I_{i\neq k} \right) \, \rho_f \, \left( \rho_{k_y} \otimes I_{i\neq k} \right) \quad \text{with}\quad        P_{k_b} = \Tr{\left[ \rho_f \left( \rho_{k_y} \otimes I_{i\neq k} \right) \right]}
      \label{eq:SI_conditioned_rho_P_bright_Ideal}
    \end{equation}    
    \begin{equation}
      \rho_{k_d} = P_{k_d}^{-1} \left( \rho_{k_{-y}} \otimes I_{i\neq k} \right) \, \rho_f \, \left( \rho_{k_{-y}} \otimes I_{i\neq k} \right) \quad \text{with}\quad P_{k_d} = \Tr{\left[ \rho_f \left( \rho_{k_{-y}} \otimes I_{i\neq k} \right) \right]}
      \label{eq:SI_conditioned_rho_P_dark_Ideal}
    \end{equation}
    \label{eq:SI_conditioned_rho_P_Ideal}
\end{subequations}

\noindent where $\rho_{k_{\pm y}}$ is the density matrix of spectator qubit $k$ in the $\pm$Y state and $I_{i\neq k}$ the identity on the Hilbert space of all remaining qubits. Using these density matrices and probabilities, we can calculate the BVL averaged over different spectator readout outcomes.\\

The analysis above assumed a perfect readout. We now include the single shot readout fidelity given by $\mathrm{F_{RO}} = \bigl( \begin{smallmatrix} F_{00} & F_{10}\\ F_{01} & F_{11}\end{smallmatrix}\bigr)$, where $F_{ij}$ is the probability to obtain outcome $j$ if the state was $i$. This provides the experimental probabilities to measure different spectator readout outcomes and their corresponding post measurement density matrices:

\begin{subequations}
    \begin{equation}
      \rho_{k_b}^e = \frac{1}{\rho_{k_b}^e} \left[ F_{00}\cdot P_{k_b}\cdot \rho_{k_b} + F_{10}\cdot P_{k_d}\cdot \rho_{k_d} \right] \quad \text{with} \quad \rho_{k_b}^e = F_{00} \cdot P_{k_b} + F_{10} \cdot P_{k_d}
      \label{eq:SI_conditioned_rho_P_bright_Experiment}
    \end{equation}    
    \begin{equation}
      \rho_{k_d}^e = \frac{1}{\rho_{k_d}^e} \left[ F_{01}\cdot P_{k_b}\cdot \rho_{k_b} + F_{11}\cdot P_{k_d}\cdot \rho_{k_d} \right] \quad \text{with} \quad \rho_{k_d}^e = F_{01} \cdot P_{k_b} + F_{11} \cdot P_{k_d}.
      \label{eq:SI_conditioned_rho_P_dark_Experiment}
    \end{equation}
    \label{eq:SI_conditioned_rho_P_Experiment}
\end{subequations}

\noindent The simulation procedure above does not include SPAM-related errors that induce dephasing on the memory qubit and other spectator qubits. As a result, the simulations always provide a BVL equal to 1 for N=0, which is not what is observed in figure \ref{Figure2}c and SI figure \ref{Fig:SI_readout_dephasing}. We compensate for this by rescaling the BVL obtained from the simulation by the fitted SPAM errors obtained from a fit through the data that the simulation targets to describe (see section \ref{sec:SI_1Spec_analytical}).\\ 

To encapsulate these SPAM-related errors in the simulations independently from experimentally obtained datasets, in the future we can target to simulate the quantum circuit used for initialization and readout and characterise the fidelities of the involved operations. A near-term improvement, that does not require additional characterisation measurements, is to include dephasing induced on nuclear-spin qubits from an electron-mediated spectator qubit readout. Namely, in our experiments, we employ a low power electron-spin readout ($\SI{100}{\pico\watt}$) for a long time ($\sim$$\SI{100}{\micro\second}$), such that we switch off the readout laser upon photon detection to prevent a subsequent electron spin excitation and corresponding potential spin-flip. As a consequence, if a spin-flip occurs before we detect a photon, there is a large uncertainty about the time at which the spin-flip happened. This relates to $F_{10}$. In the future, we can model this by assuming complete dephasing on the nuclear spins in such cases and hence replace $F_{01} \cdot P_{k_b} \cdot \rho_{k_b}$ by $F_{01} \cdot P_{k_b} \cdot \rho_{mixed}$, with $\rho_{mixed}$ the density matrix of a mixed state on the nuclear spins. Likewise we replace $F_{10} \cdot P_{k_d} \cdot \rho_{k_d}$ by $F_{10} \cdot P_{k_d} \cdot \rho_{mixed}$. \\

For the simulations corresponding to the measurement-based spectator implementation presented in the main text, we use $\mathrm{F_{RO}} = \bigl( \begin{smallmatrix} 0.88 & 0\\ 0.12 & 1\end{smallmatrix}\bigr)$ as we use a $\SI{100}{\pico\watt}$ readout laser power (see SI figure \ref{fig:SI_fig_Repump_Decay}, SI section \ref{sec:SI_Optical_Dynamics}). For the gate-based implementation, we completely omit the electron readout, and hence effectively use $\mathrm{F_{RO}} = \bigl( \begin{smallmatrix} 1 & 0\\ 0 & 1\end{smallmatrix}\bigr)$.

\section{SUPPLEMENTARY NOTE: Electron-controlled phase rotation}
\label{sec:SI_CZ}
In this section we provide the mathematical background for the electron-controlled phase rotation of the nuclear spins used in the gate-based spectator protocol (see figure \ref{Figure3}). We neglect intra-nuclear couplings ($\sim$Hz \cite{abobeih_atomic-scale_2019}) as their effect is small on the timescale of the controlled gate ($<\omega_l\approx \SI{2.3}{\micro\second}$). In that case, all nuclear spin evolutions do not depend on each other and only on the electron spin state. Following equation \eqref{eq:Hamiltonian_n_electronDependent}, we can write the electron-conditioned evolution of K nuclear spins as

\begin{align}
    H_0 &= \prod_{i=1}^K \omega_l I_{z,i} \\
    H_1 &= \prod_{i=1}^K (\omega_l - A_{\parallel,i}) I_{z,i} + A_{\perp,i} I_{x,i} \approx \prod_{i=1}^K (\omega_l - A_{\parallel,i}) I_{z,i},
    \label{eq:CZ_1}
\end{align}

\noindent where the product is a tensor product of the Hilbert spaces of the nuclear spins, and the approximation in equation \eqref{eq:CZ_1} follows from $\omega_l - A_{\parallel,i} \gg A_{\perp,i}$. Thus, each nuclear spin experiences an electron-controlled $R_Z(\theta)$ rotation if the electron is left idle. Here the control is the eigenstates of the electron (the Z basis eigenstates). In the gate-based implementation of figure \ref{Figure3}, the phase the electron picks up during the waiting time (indicated by $\delta_{ti}$ in figure \ref{Figure3}) is irrelevant and is reset by the electron spin reinitialization that follows the waiting time.

\section{SUPPLEMENTARY NOTE: Gate-based phase compensation}
\label{sec:gate-based_phase_compensation}
To alleviate electron readout induced dephasing, a gate-based implementation of the spectator approach can be employed. Here, we target to reduce the phase uncertainty of the nuclear spin qubits. After the electron is initialized to $\ket{0}_e$, we perform a spectator controlled electron flip followed by a waiting time that imprints an electron-spin dependent phase on the nuclear spins (see figure \ref{Figure3}). This section provides insight in the mathematical background underlying this gate based approach. We first derive the gate-based effect considering a pure state ($P(\theta) = \delta(\theta)$) of the nuclear spin qubits in the XY plane of the Bloch sphere, and use that result to generalize for a nuclear spin qubit phase probability density function.\\

Given a phase $\theta$ on the memory and a phase correlation between spectator qubit $i$ and the memory given by $\theta_i = g_i \cdot \theta$ we write 

\begin{equation}
    \ket{\psi} = \prod_{(i\neq k)} \left( \frac{1}{\sqrt{2}}\left[ \ket{0}_i +e^{i g_i \theta}\ket{1}_i \right] \right) \otimes \frac{1}{\sqrt{2}}\left[ \ket{0}_k +e^{i g_k \theta}\ket{1}_k \right] \otimes \frac{1}{\sqrt{2}}\left[ \ket{0}_m +e^{i \theta}\ket{1}_m \right] \otimes \ket{0}_e,
    \label{eq:SI_gate_based_psi_start}
\end{equation}

\noindent where the indices refer to the memory qubit ($m$) or the spectator qubits ($i$ \& $k$). SI section \ref{sec:RObasis_phi_c} showed that for a normally distributed phase probability density function centered at $\theta = 0$, the optimal spectator readout angle is $\pm \frac{\pi}{2}$. Hence, here we perform a spectator $k$ controlled electron flip where we flip the electron if the spectator is in $\ket{-i}$. This modifies $\ket{\psi}$ to

\begin{align}
    \begin{split}
        \ket{\psi} = & \prod_{(i\neq k)} \left( \frac{1}{\sqrt{2}}\left[ \ket{0}_i +e^{i g_i \theta}\ket{1}_i \right] \right) \otimes \frac{1}{\sqrt{2}}\left[ \ket{0}_m +e^{i \theta}\ket{1}_m \right] \\
        &\otimes \frac{1}{2\sqrt{2}}\left( 1-ie^{ig_k\theta} \right)\left[ \ket{0}_k + i\ket{1}_k \right] \otimes \ket{0}_e + \\
        & \prod_{(i\neq k)} \left( \frac{1}{\sqrt{2}}\left[ \ket{0}_i +e^{i g_i \theta}\ket{1}_i \right] \right) \otimes \frac{1}{\sqrt{2}}\left[ \ket{0}_m +e^{i \theta}\ket{1}_m \right] \\
        &\otimes \frac{1}{2\sqrt{2}}\left( 1+ie^{ig_k\theta} \right)\left[ \ket{0}_k - i\ket{1}_k \right] \otimes \ket{1}_e.
        \label{eq:SI_gate_based_psi_post_CNOT}
    \end{split}
\end{align}

\noindent In the rotating frame of the nuclear spins with the electron in the $\ket{0}$ state, a subsequent waiting time $t_w$ imprints a phase $g_i \phi_c = g_i (\omega_{1_m} - \omega_0)t_w$ on the nuclear spin qubits correlated with the electron $\ket{1}$ state:

\begin{align}
    \begin{split}
        \ket{\psi}(\phi_c) = & \prod_{(i\neq k)} \left( \frac{1}{\sqrt{2}}\left[ \ket{0}_i +e^{i g_i \theta}\ket{1}_i \right] \right) \otimes \frac{1}{\sqrt{2}}\left[ \ket{0}_m +e^{i \theta}\ket{1}_m \right] \\
        &\otimes \frac{1}{2\sqrt{2}}\left( 1-ie^{ig_k\theta} \right)\left[ \ket{0}_k + i\ket{1}_k \right] \otimes \ket{0}_e \\
        & + \prod_{(i\neq k)} \left( \frac{1}{\sqrt{2}}\left[ \ket{0}_i +e^{i g_i (\theta+\phi_c)}\ket{1}_i \right] \right) \otimes \frac{1}{\sqrt{2}}\left[ \ket{0}_m +e^{i (\theta+\phi_c)}\ket{1}_m \right] \\
        &\otimes \frac{1}{2\sqrt{2}}\left( 1+ie^{ig_k\theta} \right)\left[ \ket{0}_k - ie^{i g_k \phi_c}\ket{1}_k \right] \otimes \ket{1}_e.
        \label{eq:SI_gate_based_psi_post_wait}
    \end{split}
\end{align}

\noindent The subsequent electron reinitialization mathematically translates to tracing out the electron state of equation \eqref{eq:SI_gate_based_psi_post_wait}. This leaves the nuclear spins in state $\ket{\psi}_0$ with probability $p_0$ and in state $\ket{\psi}_1$ with probability $p_1$, with 

\begin{subequations}
    \begin{align}
        \begin{split}
          \ket{\psi}_0 &= \prod_{(i\neq k)} \left( \frac{1}{\sqrt{2}}\left[ \ket{0}_i +e^{i g_i \theta}\ket{1}_i \right] \right) \otimes \frac{1}{\sqrt{2}}\left[ \ket{0}_m +e^{i \theta}\ket{1}_m \right] \\
          &\otimes \frac{1}{2\sqrt{2}}\left( 1-ie^{ig_k\theta} \right)\left[ \ket{0}_k + i\ket{1}_k \right] \\
          p_0 &= \frac{1}{2} \left[ 1 + \sin{(g_k \theta)} \right]
        \end{split}
    \end{align}    
    \begin{align}
        \begin{split}
          \ket{\psi}_1 &= \prod_{(i\neq k)} \left( \frac{1}{\sqrt{2}}\left[ \ket{0}_i +e^{i g_i (\theta+\phi_c)}\ket{1}_i \right] \right) \otimes \frac{1}{\sqrt{2}}\left[ \ket{0}_m +e^{i (\theta+\phi_c)}\ket{1}_m \right] \\
          &\otimes \frac{1}{2\sqrt{2}}\left( 1+ie^{ig_k\theta} \right)\left[ \ket{0}_k - ie^{i g_k \phi_c}\ket{1}_k \right] \\
          p_1 &= \frac{1}{2} \left[ 1 - \sin{(g_k \theta)} \right].
        \end{split}
    \end{align} 
    \label{eq:SI_gate_based_psi_post_eTrace}
\end{subequations}

Equations \eqref{eq:SI_gate_based_psi_post_eTrace} show two important observations: 1) there is no phase correlation between the state of spectator $k$ and the other qubits, which corroborates that the gate-based implementation effectively implements a measurement on the phase of spectator $k$. 2) the post gate-based approach probability density function of the phase $\theta$ is $\frac{1}{2}[1+\sin{(g_k\theta)}]\delta(\theta) + \frac{1}{2}[1-\sin{(g_k\theta)}]\delta(\theta-\phi_c)$. For a general initial phase distribution $P(\theta)$, the posterior phase distribution $P(\theta|k_{-})$ after the gate-based spectator approach is given by equation \eqref{eq:SI_gate-based_posterior}. Here $k_{-}$ indicates that an electron flip was executed for the state correlated to the -Y state of spectator $k$. Execution of the electron flip correlated to the +Y state of spectator $k$ is equivalent to replacing $\phi_c$ for $-\phi_c$ in equation \eqref{eq:SI_gate-based_posterior}:

\begin{equation}
    P(\theta,\phi_c|k_{-}) = \frac{1}{2}\left[1 + \sin{(g_k \theta)} \right] P(\theta) + \frac{1}{2}\left[1 - \sin{(g_k \theta)}\right] P(\theta-\phi_c).
    \label{eq:SI_gate-based_posterior}
\end{equation}

\begin{figure*}[!ht]
\includegraphics[width=\textwidth]{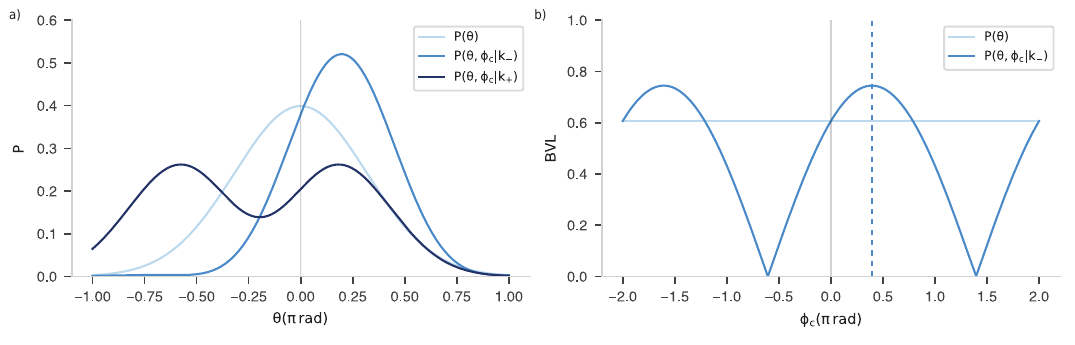}
\caption{\label{fig:SI_Gate-based} \textbf{Gate based spectator implementation.} \textbf{a)} Initial normally distributed PDF $\mathrm{P(\theta)}$ with standard deviation $\sigma=1$ (light blue). The posterior PDF if an electron flip is correlated with the -Y state on spectator $k$: $\mathrm{P(\theta,\phi_c|k_{-})}$ (blue) and when correlated with the +Y state $\mathrm{P(\theta,\phi_c|k_{+})}$ (dark blue). $\phi_c$ in this figure is set to the dashed line in b), which maximizes the BVL corresponding to $\mathrm{P(\theta,\phi_c|k_{-}))}$. \textbf{b)} BVL for the prior ($\mathrm{P(\theta)}$, light blue) and posterior ($\mathrm{P(\theta,\phi_c|k_{-}))}$, blue) PDF as a function of $\phi_c$.}
\end{figure*}

In SI figure \ref{fig:SI_Gate-based}a, $P(\theta,\phi_c|k_{-})$ and $P(\theta,\phi_c|k_{+})$ are depicted for a normal distribution of $P(\theta)$ with $\sigma=1$ and $\phi_c=0.34$. In line with figure \ref{Figure3}, to narrow the phase distribution, it is crucial to choose the right spectator state (+Y or -Y) based on which an electron flip is executed. Note that experimentally only the magnitude of $\phi_c$ can be chosen. Namely $t_w\geq0$ and the sign is determined by $\omega_{1}-\omega_{0}$. The narrowing of $P(\theta,\phi_c|k_{-})$ with respect to $P(\theta)$ is reflected in an increased BVL as demonstrated by SI figure \ref{fig:SI_Gate-based}b. The blue dashed line reflects the optimal angle $\phi_c$ for which the plots in SI figure \ref{fig:SI_Gate-based}a are generated. The $\phi_c$ values for which the BVL is zero correspond to the situation shown in the right bottom of figure \ref{Figure3}b. Increasing $\phi_c$ will move the BVL through a discontinuity at 0. This discontinuity is explained as upon the zero crossing the Bloch vector obtains a $\pi$ phase shift.

\section{SUPPLEMENTARY NOTE: Optical dynamics in spectator qubit measurements}
\label{sec:SI_Optical_Dynamics}

\begin{figure*}[!ht]
\includegraphics[width=\textwidth]{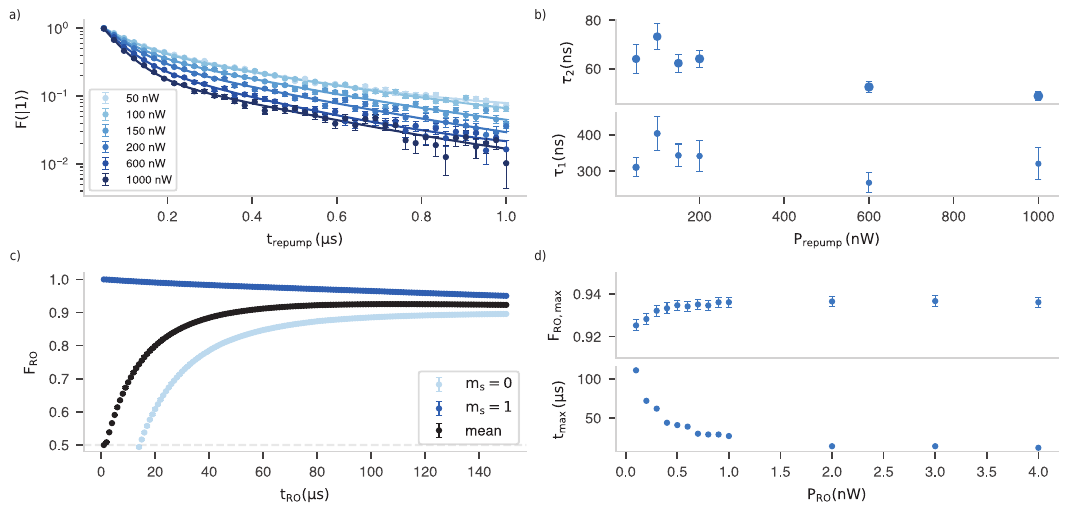}
\caption{\label{fig:SI_fig_Repump_Decay} \textbf{Electron optical dynamics.} \textbf{a)} Depletion of the population in $\ket{1}$ under constant illumination of the repump laser for different a laser powers. Fitted line is $F(\ket{1}) = A_1\exp{(-\frac{t}{\tau_1})} +A_2\exp{(-\frac{t}{\tau_2})} + a$. \textbf{b)} Fitparameters for the curves in (a). The size of the markers reflects $\frac{A_1}{A_1+A_2}$ (bottom) $\left( \frac{A_2}{A_1+A_2} \right)$ (top). \textbf{c)} Exemplary single-shot readout curve. The applied readout power here is $\SI{100}{\pico\watt}$. The light (dark) blue curve indicates the probability to (not) have detected a photon after applying the RO laser for a time $\mathrm{t_{RO}}$ if the initial state is $\ket{0}$ ($\ket{1}$). Black is the average readout fidelity, averaged over the light and dark blue data. \textbf{d)} Maximum readout fidelities and corresponding $\mathrm{t_{RO}}$ at which these maxima are achieved for different readout powers. The readout power used in the spectator measurements in this work is $\SI{100}{\pico\watt}$. Upon photon detection, we turn off the readout laser, and otherwise terminate the readout after $\SI{98}{\micro\second}$.}
\end{figure*}

In this section, we first provide insight in the powers and duration used for the electron repump during the entanglement generation sequence. Secondly, we do this for the power and duration involved in the spectator readout. In SI figure \ref{fig:SI_fig_Repump_Decay}a, we plot the fidelity of the electron spin with respect to the $\ket{1}$ state after having the electron spin initialized in the $\ket{1}$ state and applying a resonant repump laser for a time $t_{\mathrm{repump}}$ for different powers. We fit a double exponential decay (see figure caption) and plot the extracted timescales in SI figure \ref{fig:SI_fig_Repump_Decay}b. The longer the decay timescales, the longer the electron undergoes continuous optical cycles and the longer the electron experiences a stochastic evolution. In the experimental results in this work, we set the repump power to $\SI{600}{\nano\watt}$ and the repump time to $\sim$$\SI{500}{\nano\second}$. A longer spin-pump time would improve the electron reinitialization fidelity, at the cost of a longer sequence time. Infidelities in the electron spin reinitialization act as correlated noise, which the spectator qubit approach can mitigate. \\

In the spectator qubit readout, we want to simultaneously optimize the probability to obtain the right readout outcome and the probability that the post-measurement state is equal to the measured state. We read out by selectively addressing the $\ket{0}$ electron state \cite{robledo_high-fidelity_2011}. In case the electron is in the $\ket{0}$ state, the electron cycles through the excited state, thereby emitting photons. Photon detection measures the electron spin state $\ket{0}$. In this process, spin-flips in the excited state can change the electron spin state. To minimize such spin flips, we avoid unnecessary excitations by using a weak laser power, such that a feedback signal ($\SI{1}{\micro\second}$ clock cycle) can be used to rapidly turn off the laser upon detection of a photon \cite{cramer_repeated_2016,abobeih_fault-tolerant_2022}. \\

In SI figure \ref{fig:SI_fig_Repump_Decay}c, we show the probability to (not) measure a photon in case the electron was prepared in the $\ket{0}$ ($\ket{1}$) state upon application of a $\SI{100}{\pico\watt}$ laser pulse for duration $t_{\mathrm{RO}}$. From this, we can extract what the maximum achievable fidelity averaged over the $\ket{0}$ and $\ket{1}$ is, and at what $t_{\mathrm{RO}}$ we obtain this fidelity. We do this for different laser powers and plot these numbers in SI figure \ref{fig:SI_fig_Repump_Decay}d. For lower repump powers, we barely observe a drop in the maximum attainable fidelity. Therefore we pick a low readout power in our experiments ($\SI{100}{\pico\watt}$) to benefit the ability to turn off the readout laser upon photon detection. \\

\section{SUPPLEMENTARY NOTE: Extended datasets}
\label{sec:SI_Extended_datasets}

\begin{figure*}[!ht]
\includegraphics[width=\textwidth]{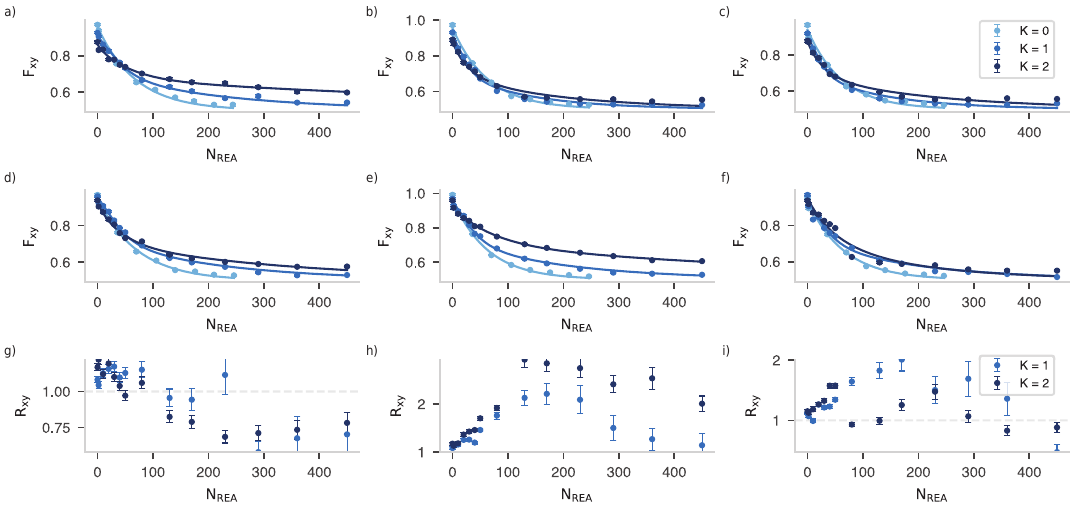}
\caption{\label{fig:SI_fig_7_3_datasets_xy} \textbf{Three datasets of spectator data}. Each column shows data of the same experimental sequence, and the time separation of data acquisition is at least one week. During all datasets identical magnetic field stabilisation measurements were ran. The first row shows the spectator effect in the measurement-based protocol, the second row in the gate-based protocol, and the third row shows the ratio $R_{xy}$ between these protocols as explained in the main text.}
\end{figure*}

\begin{figure*}[!ht]
\includegraphics[width=\textwidth]{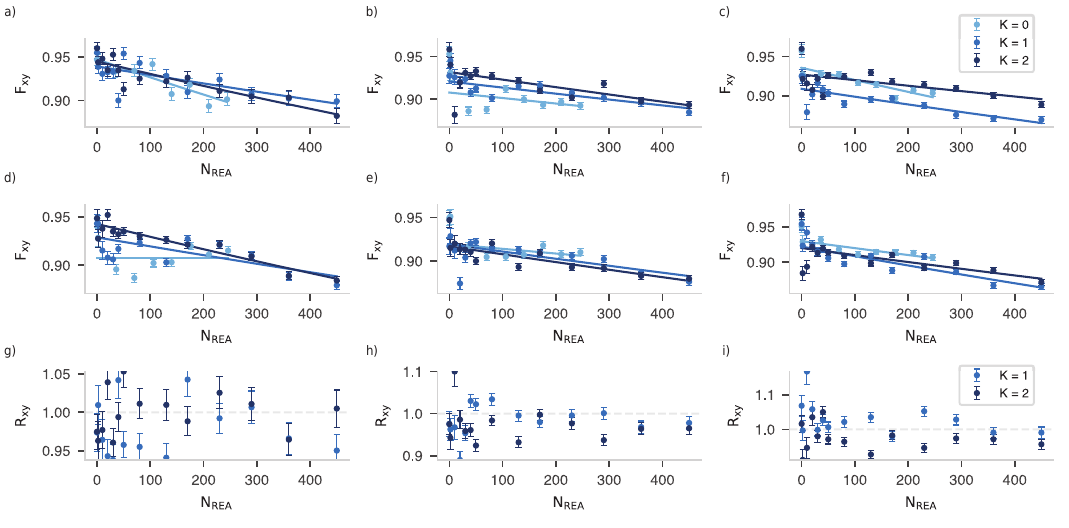}
\caption{\label{fig:SI_fig_8_3datasets_z} \textbf{Same as SI figure \ref{fig:SI_fig_7_3_datasets_xy}, but here with the memory qubit initialized in the Z-basis}}
\end{figure*}

Here we show all three datasets obtained for the selected spectator and memory qubit(s) as used in the main text. The datasets used the same experimental sequence and environmental stabilisation protocols, and were taking with at least a week time difference. In SI figure \ref{fig:SI_fig_7_3_datasets_xy}, we show the data averaged for the memory qubit initialized in the X- and Y- basis. In SI figure \ref{fig:SI_fig_8_3datasets_z}, we show the data for the memory qubit initialized in the Z-basis. In both figures, panel 'a)` is shown in figure \ref{Figure2} in the main text. The general trends as described in the main text are also reflected by the other datasets.

\FloatBarrier

\section{SUPPLEMENTARY NOTE: Alternative spectator qubit assignments}
\label{sec:SI_alternative_spectators}
Here we show the complete underlying data on which figure \ref{Figure3}d is based. In each row we use a different strategy in terms of which and how many carbon spins to use as spectator qubits, and in each column we use a different carbon spin as memory qubit. This is reflected in the title of each panel, see figure caption for more details.

\begin{figure*}[!ht]
\includegraphics[width=\textwidth]{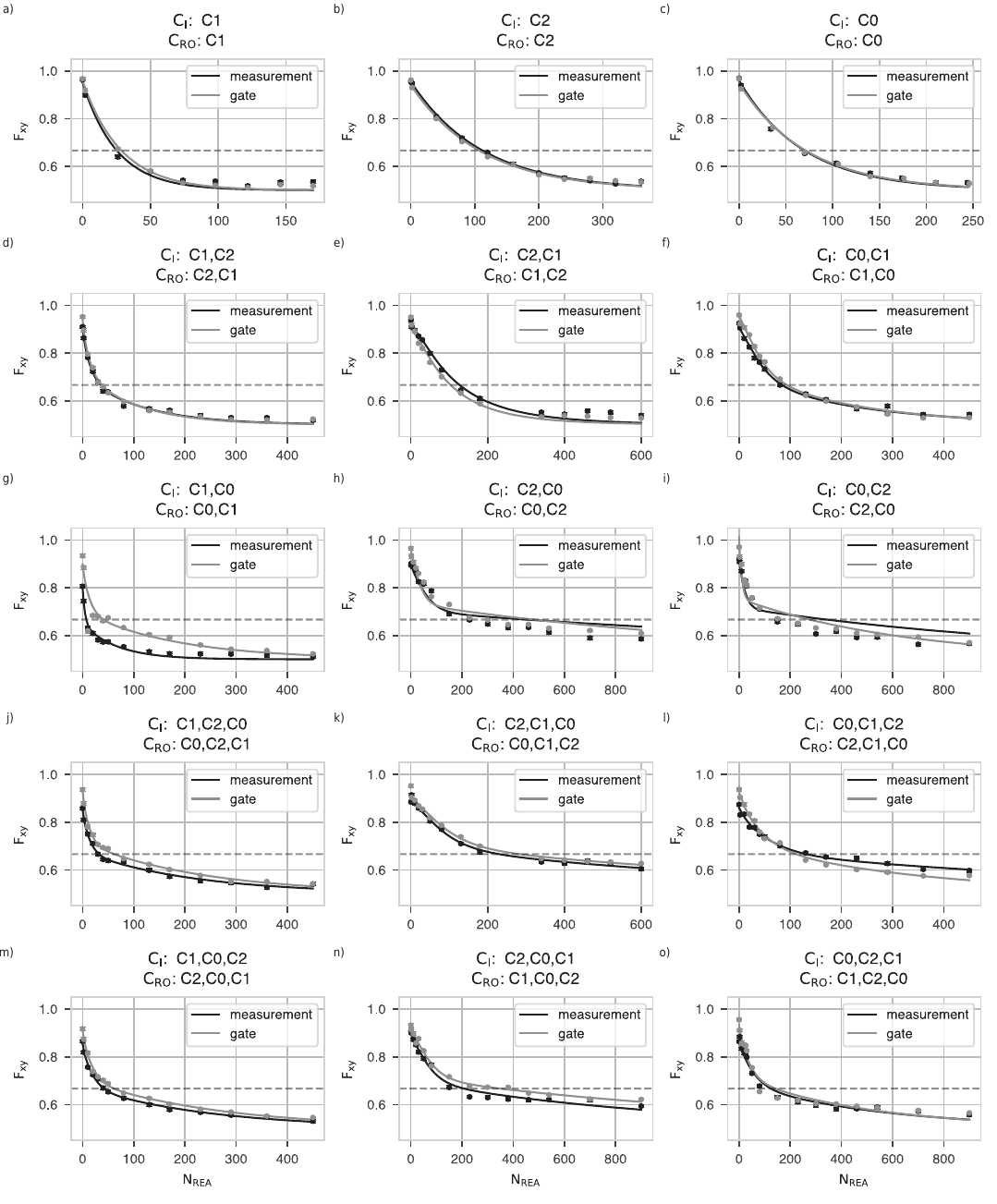}
\caption{\label{fig:SI_fig_9_alternative_spectators} \textbf{Comparison of measurement- and gate-based protocols of alternative assignment of spectator and memory qubit.} The labels on the top indicate which qubits are function as spectator and  memory qubit. $\mathrm{C_I}$:\, $\mathrm{C_a}, \, \mathrm{C_b}, \, \mathrm{C_c}$ means $\mathrm{C_a}$ is the memory qubit, $\mathrm{C_b}$ is the first spectator qubit and $\mathrm{C_c}$ is the second spectator qubit. $\mathrm{C_R}$ indicates the order in which the qubits are read out (measurement-based protocol) or mapped to the electron spin with a consequent waiting time (gate-based protocol). Data is averaged over data with the memory qubit initialized in the X- and Y-basis.}
\end{figure*}

\begin{figure*}[!ht]
\includegraphics[width=\textwidth]{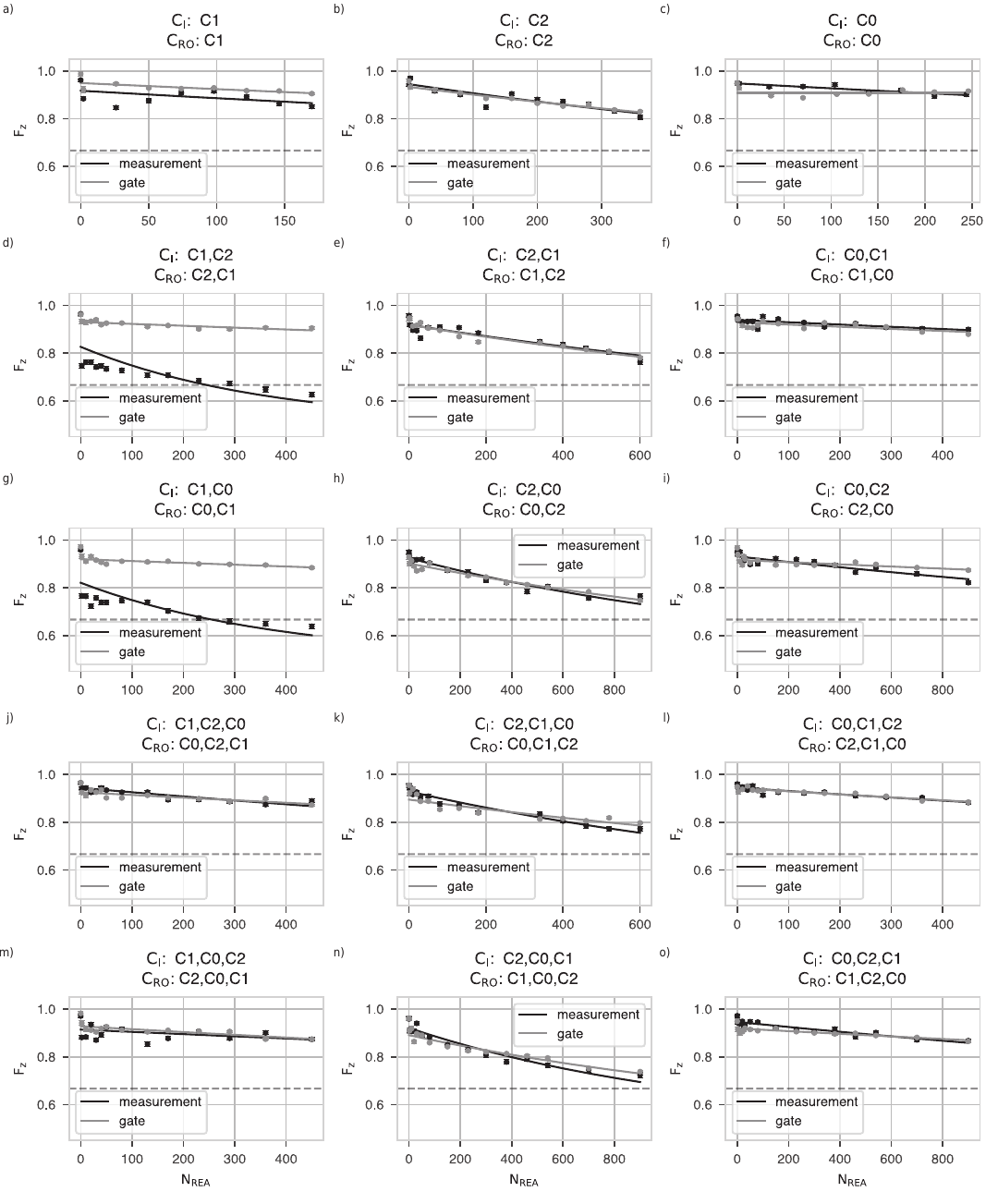}
\caption{\label{fig:SI_fig_10_alternative_spectators} \textbf{Same as SI figure \ref{fig:SI_fig_9_alternative_spectators}, but here with the memory qubit initialized in the Z-basis}}
\end{figure*}

\FloatBarrier

\end{document}